\documentclass[12pt,a4paper]{article}
\usepackage{xcolor}
\usepackage{jheppub}
\usepackage{epstopdf}
\usepackage{graphicx}
\usepackage{epsfig}
\usepackage{dcolumn}  
\usepackage{bm}    
\usepackage{amssymb} 
\usepackage{amsmath,bm}
\usepackage{amsfonts}    
\usepackage{slashed}  
\usepackage{youngtab}
\usepackage[mathscr]{euscript}
\usepackage{epsfig}
\usepackage{verbatim}
\usepackage{tensor}

\newdimen\tableauside\tableauside=1.0ex
\newdimen\tableaurule\tableaurule=0.4pt
\newdimen\tableaustep
\def\phantomhrule#1{\hbox{\vbox to0pt{\hrule height\tableaurule width#1\vss}}}
\def\phantomvrule#1{\vbox{\hbox to0pt{\vrule width\tableaurule height#1\hss}}}
\def\sqr{\vbox{%
		\phantomhrule\tableaustep
		\hbox{\phantomvrule\tableaustep\kern\tableaustep\phantomvrule\tableaustep}%
		\hbox{\vbox{\phantomhrule\tableauside}\kern-\tableaurule}}}
\def\squares#1{\hbox{\count0=#1\noindent\loop\sqr
		\advance\count0 by-1 \ifnum\count0>0\repeat}}
\def\tableau#1{\vcenter{\offinterlineskip
		\tableaustep=\tableauside\advance\tableaustep by-\tableaurule
		\kern\normallineskip\hbox
		{\kern\normallineskip\vbox
			{\gettableau#1 0 }%
			\kern\normallineskip\kern\tableaurule}%
		\kern\normallineskip\kern\tableaurule}}
\def\gettableau#1 {\ifnum#1=0\let\next=\null\else
	\squares{#1}\let\next=\gettableau\fi\next}

\allowdisplaybreaks[1]

\newcommand{\be}{ \begin{equation}}
\newcommand{\ee}{\end{equation}}
\newcommand{\bea}[1]{\begin{eqnarray}\label{#1} }
\newcommand{\eea}{\end{eqnarray}}

\def\ZZZ{{\hskip-3pt\hbox{ Z\kern-1.6mm Z}}}
\def\zzz{{\hskip-3pt\hbox{ z\kern-1mm z}}}

\newcommand{\abs}[1]{\left| #1 \right|}

\def\one{{\hbox{ 1\kern-.8mm l}}}
\def\zero{{\hbox{ 0\kern-1.5mm 0}}}

\title{BPS spectrum on AdS$_\mathbf{3}\times $S$^\mathbf{3} \times $S$^\mathbf{3} \times $S$^\mathbf{1}$}

\author{Lorenz Eberhardt$^a$, Matthias R.\ Gaberdiel$^a$, Rajesh Gopakumar$^b$,  and Wei Li$^c$} 
\affiliation{$^a$ Institut f\"ur Theoretische Physik, ETH Zurich, \\
\hspace*{0.3cm}CH-8093 Z\"urich, Switzerland}
\affiliation{$^b$ International Centre for Theoretical Sciences-TIFR, \\
\hspace*{0.3cm}Survey No. 151, Shivakote, Hesaraghatta Hobli, \\
\hspace*{0.3cm}Bengaluru North, India 560 089}
\affiliation{$^c$ Institute of Theoretical Physics, Chinese Academy of Science\\
\hspace*{0.3cm} 100190 Beijing, P.R.\ China}
\emailAdd{eberhardtl@itp.phys.ethz.ch, gaberdiel@itp.phys.ethz.ch, \\ rajesh.gopakumar@icts.res.in, weili@itp.ac.cn}

\abstract{The BPS spectrum of string theory on AdS$_3\times {\rm S}^3 \times {\rm S}^3 \times {\rm S}^1$
is determined using a world-sheet description in terms of WZW models. It is found that the theory
only has BPS states with $j^+ = j^-$ where $j^{\pm}$ refer to the spins of the $\mathfrak{su}(2)$ algebras of the 
large ${\cal N}=4$ superconformal algebra. We then re-examine the BPS spectrum of the corresponding
supergravity and find that, in contradistinction to previous claims in the literature, also in supergravity only the states
with $j^+=j^-$ are BPS. This resolves a number of long-standing puzzles regarding the BPS spectrum of string theory
and supergravity in this background. 
}

\begin{document}

\maketitle

\makeatletter
\g@addto@macro\bfseries{\boldmath}
\makeatother
\section{Introduction}

For the case of AdS$_3$, the AdS/CFT correspondence has been understood in quite some detail. In particular,
there is very convincing evidence that the CFT dual of string theory on AdS$_3\times {\rm S}^3 \times {\cal M}_4$ 
is (on the moduli space of) the symmetric orbifold of ${\cal M}_4$, where ${\cal M}_4=\mathbb{T}^4$ or 
${\cal M}_4 = {\rm K3}$, see e.g.\ \cite{David:2002wn} for a review. For example, the BPS
spectrum of the two descriptions matches perfectly, and their correlation functions agree 
\cite{Gaberdiel:2007vu,Dabholkar:2007ey,deBoer:2008ss}. 
More recently, it was also found that the symmetric orbifold
is a natural extension of the CFT dual of a supersymmetric higher spin theory on AdS$_3$ \cite{Gaberdiel:2014cha}. For
either choice of ${\cal M}_4$,  the dual CFT has the (small) ${\cal N}=4$ superconformal symmetry 
\cite{Ademollo:1976wv,Eguchi:1987sm,Eguchi:1987wf}. 

On the other hand, the situation is much less clear \cite{Gukov:2004ym} for the other maximally supersymmetric  AdS$_3$ background
AdS$_3\times {\rm S}^3 \times {\rm S}^3 \times {\rm S}^1$, and no convincing proposal for what precisely the dual CFT should be 
exists to date, see however \cite{Tong:2014yna} for a recent attempt.  This is a bit
surprising since the corresponding dual CFT has an even larger symmetry algebra, the so-called large ${\cal N}=4$
superconformal algebra $A_\gamma$ \cite{Sevrin:1988ew}, see also \cite{Schoutens:1988ig,Spindel:1988sr,VanProeyen:1989me,Sevrin:1989ce}.
One of the reasons why the case with large ${{\cal N}=4}$ superconformal symmetry is more difficult is related
to the fact that the BPS bound for $A_\gamma$ is in general stronger than the corresponding BPS bound 
\cite{Gunaydin:1988re,Petersen:1989zz,Petersen:1989pp} of  the supergravity symmetry algebra $D(2,1|\alpha)$
\cite{de Boer:1999rh}. Both algebras contain an $\mathfrak{su}(2)\oplus \mathfrak{su}(2)$
subalgebra, and the BPS bounds for the two algebras only agree if the spins with respect to these two algebras $(j^+,j^-)$ coincide, 
$j^+=j^-$.  In particular, this leads to the somewhat mysterious phenomenon that any supergravity BPS state with $j^+\neq j^-$
{\em has to} acquire non-trivial quantum corrections upon quantisation in order to satisfy (let alone saturate)
the BPS bound of $A_\gamma$ \cite{de Boer:1999rh,Gukov:2004ym}.  This is not just a theoretical
possibility since, according to the analysis  of \cite{de Boer:1999rh}, such BPS states exist in supergravity.
\smallskip

In this paper we revisit this somewhat unsatisfactory situation. We begin by studying string theory on 
AdS$_3\times {\rm S}^3 \times {\rm S}^3 \times {\rm S}^1$
from a world-sheet prespective, following  the analysis of \cite{Elitzur:1998mm}. The relevant background has pure NS-NS
flux, and the AdS$_3$ factor is described by a WZW model associated to $\mathfrak{sl}(2,\mathbb{R})$ as in \cite{Maldacena:2000hw},
while for the ${\rm S}^3$ factors we have the familiar $\mathfrak{su}(2)$ WZW models. It was shown in \cite{Elitzur:1998mm} that
the spacetime CFT (i.e., the dual CFT) has large ${\cal N}=4$ superconformal symmetry $A_\gamma$. Thus we can analyse which
states of the spacetime spectrum saturate the $A_\gamma$ BPS bound, and we find that the only states with this property
have $j^+=j^-$. Furthermore, since string theory reduces to supergravity in the limit of vanishing string size, our analysis also 
leads to a prediction for what the supergravity BPS spectrum should be. This suggests that all the BPS states of 
supergravity also have $j^+=j^-$. 

Given that this conclusion is in contradiction to the claim of \cite{de Boer:1999rh}, we perform then a first principle supergravity
analysis, following the strategy of \cite{Deger:1998nm} suitably adjusted to the current setting. We should stress that this somewhat
tedious analysis was not done in \cite{de Boer:1999rh} where it was simply assumed that for each harmonic there would be a 
BPS state --- as was indeed the case for AdS$_3\times {\rm S}^3 \times {\cal M}_4$ with ${\cal M}_4=\mathbb{T}^4$ or 
${\rm K3}$. Using this assumption the BPS states were then organised into supermultiplets, using group theoretic methods
\cite{de Boer:1999rh}.

In our analysis,  we start with $9$-dimensional supergravity (with a pure NS-NS background) and compactify
it on ${\rm S}^3 \times {\rm S}^3$, making an expansion in terms of spherical harmonics on the two spheres. We will concentrate on the 
scalar fields coming from the NS-NS fields in $9$ dimensions; this is sufficient for the analysis of the BPS spectrum since 
every BPS multiplet in the list of  \cite{de Boer:1999rh} contains at least one such field.  The resulting field equations
are then Klein-Gordon equations from the viewpoint of AdS$_3$, and hence we can easily read off their masses 
as a function of the spins along the two ${\rm S}^3$'s. The analysis is however quite tedious, but with the help of {\tt Mathematica},\footnote{For
the convenience of the reader we have appended the {\tt Mathematica} workbook as an ancillary file to the {\tt arXiv} submission.}
we have managed to find all eigenfunctions and identify their corresponding masses. The result turns out to be exactly as
predicted from the world-sheet analysis: the only BPS states of supergravity appear for $j^+=j^-$.
\smallskip

This is the main result of this paper. As was alluded to before, it resolves the puzzle about the mysterious 
quantum corrections of supergravity BPS states: the only BPS states that would have to behave in this 
manner arise for $j^+\neq j^-$ --- and the resolution is simply that no such supergravity BPS states exist! 
In fact, our analysis also shows that the actual masses of the supergravity states with $j^+\neq j^-$ do not
just obey the supergravity BPS bound, but in fact also the stronger $A_\gamma$ BPS bound (and saturate neither). 

As a consequence of our analysis, also the question about the CFT dual to string theory on 
AdS$_3\times {\rm S}^3 \times {\rm S}^3 \times {\rm S}^1$ needs to be re-examined. In particular,
the most natural candidate theory at least for the case when the two ${\rm S}^3$'s have the same 
size, the symmetric orbifold of the so-called ${\cal S}_0$ theory \cite{Elitzur:1998mm,Gukov:2004ym}, was
largely discarded in \cite{Gukov:2004ym} because of its failure to reproduce the BPS spectrum of supergravity --- but
since the latter was wrongly identified, this conclusion does not hold any more. In fact, it now seems that 
this symmetric orbifold is a viable candidate, and we are in the process of exploring this possibility further \cite{inprep}.
\medskip

The paper is organised as follows. In Section~\ref{sec:worldsheet} we review the world-sheet description of this 
background and then analyse its spacetime BPS spectrum. The main result of this section is eq.~(\ref{main}) which
gives a lower bound on the conformal dimension of any spacetime state as a function of its spins. It follows
from this bound that the only BPS states arise for $j^+=j^-$. We furthermore speculate in Section~\ref{sec:sugraint}
what the supergravity incarnation of this result should be. In Section~\ref{sec:sugra} we then perform 
the supergravity analysis from first principles: we first determine the $9$-dimensional vacuum solution 
(Section~\ref{sec:3.1}), and deduce the equations of motion for the fluctuations around this vacuum solution
(Section~\ref{sec:3.2}).  The various components are then expanded in terms of 
spherical harmonics (Section~\ref{sec:3.3}), and the different modes are diagonalised 
into eigenfunctions of the Laplace operator on AdS$_3$; for three of the scalar fields
this is done explicitly in Section~\ref{sec:decoupled}, while the analysis of the remaining seven scalars is quite
complicated and has been relegated to an appendix (Appendix~\ref{app:sugra}), with only the results being given in 
Section~\ref{sec:remain}. The full scalar spectrum is then analysed in Section~\ref{sec:full}, and the above
statements about the supergravity BPS bound are derived. Our analysis culminates in the description of the 
full supergravity spectrum in eq.~(\ref{multiplet_spectrum}). Finally, Section~\ref{sec:concl} contains our
conclusions and outlines future directions of research. There are three appendices: in
Appendix~\ref{app:superalgebra} we review the $D(2,1|\alpha)$ and the large ${\cal N}=4$ superconformal 
algebra $A_\gamma$ and describe their BPS bounds. Appendix~\ref{app:sugra}  contains part of the
general supergravity analysis, while Appendix~\ref{app:special} deals with the special features that arise
for the harmonics with small spin.

\section{The world-sheet analysis}\label{sec:worldsheet}

In this section we analyse the string spectrum on  AdS$_3\times {\rm S}^3 \times {\rm S}^3 \times {\rm S}^1$ for the 
case of pure NS-NS flux. This background may be described by a WZW model as in \cite{Elitzur:1998mm}.
More specifically, the AdS$_3$ factor is captured by an ${\cal N}=1$ superconformal $\mathfrak{sl}(2,\mathbb{R})$ 
WZW model at level $k$, while for the two ${\rm S}^3$ factors we have ${\cal N}=1$ superconformal WZW models based 
on $\mathfrak{su}(2)$ at levels $k^\pm$. Finally, the ${\rm S}^1$  factor is just described by  a free boson and a free fermion. 
After decoupling the respective fermions, the bosonic currents (that commute with the fermions) 
then have levels $k+2$, and $k^\pm-2$, respectively. The requirement that the string theory is critical, i.e.,
has total central charge $c=15$, implies a relation between the levels, see e.g., \cite{Elitzur:1998mm}
\be\label{krel}
\frac{1}{k} = \frac{1}{k^+} + \frac{1}{k^-} \ , \ \ \hbox{i.e.} \qquad 
k = \frac{k^+ k^-}{k^+ + k^-} \ . 
\ee
Geometrically, $k^\pm$ describe the sizes of the two ${\rm S}^3$'s, and (\ref{krel}) implies that the size of the 
AdS$_3$ space (that is described by $k$) is fixed in terms of these two. 

To be more specific, we denote the bosonic modes of $\mathfrak{sl}(2,\mathbb{R})$ at level $k+2$ by 
$J^a_n$, while $K^{\pm,a}_n$ are the modes for $\mathfrak{su}(2)$ at level $k^{\pm}-2$. (In either case  $a=3,\pm$.) 
The corresponding
fermions will be denoted by $\psi^a_r$ and $\chi^{\pm,a}_r$, respectively, while for the ${\rm S}^1$ factor we have 
the bosonic and fermionic modes $\alpha_n$ and $b_r$, respectively. We also denote the corresponding
supersymmetric currents by $\mathcal{J}^a_n$ and $\mathcal{K}^{\pm,a}_n$; their levels are then $k$, and $k^\pm$, 
respectively. The ${\cal N}=1$ superconformal generators can be constructed in the usual manner
\cite{DiVecchia:1984nyg}.

In the following we shall mainly concentrate on the NS sector of the theory; we shall
come back to the R sector at the end of this section. There the physical states $\Phi$
are characterised by the condition
\begin{align}\label{physstate}
L_n \Phi=0\ ,\ \ n>0\ ,\qquad G_r \Phi=0\ ,\ \ r>0\ ,\qquad \left(L_0-\frac{1}{2}\right) \Phi=0\ ,
\end{align}
with a similar relation for the right-movers. We shall furthermore mainly consider the unflowed sector 
of the $\mathfrak{sl}(2,\mathbb{R})$ WZW model \cite{Maldacena:2000hw}; we will later comment on the 
situation in the flowed sectors. 

The affine representations that appear in the spectrum are (in the unflowed sector) 
conventional highest weight representations, and can be characterised by 
$(j_0;j_0^+,j_0^-)$, where
$j_0$ is the spin of the  $\mathfrak{sl}(2,\mathbb{R})$ highest weight representation, while
$j_0^\pm$ are the spins of the two $\mathfrak{su}(2)$ highest weight representations. In our conventions, 
$j_0\geq 0$ labels the `highest weight state' that is characterised by 
\be
J^-_0 \, |j_0\rangle = J^a_n \, |j_0 \rangle = 0 \ , \ \ n>0 \ , \quad \hbox{and} \quad J^3_0 \, |j_0\rangle = j_0 \, |j_0\rangle \ ,
\ee
and the Casimirs of the $\mathfrak{sl}(2,\mathbb{R})$ and $\mathfrak{su}(2)$ representations are 
\be
C^{{\rm SL}(2)} = - j_0 (j_0 - 1) \ , \qquad C^{{\rm SU}(2)} = j_0^\pm (j_0^\pm + 1) \ . 
\ee
Let us denote by $N$ the excitation number of the physical state; then
the mass-shell condition -- the last equation in (\ref{physstate}) --- implies that 
\be
N = \frac{1}{2} + \frac{j_0 (j_0 -1)}{k} - \frac{j_0^+ (j_0^+ + 1)}{k^+} - \frac{j_0^- (j_0^- + 1)}{k^-} \ .
\ee
We have analysed the other two constraints of (\ref{physstate}) on the low-lying ($N\leq \frac{3}{2}$) 
physical states following the analysis of \cite{Evans:1998qu}. Modulo 
spurious null-states, we have found that 
the resulting spectrum has the standard form one expects from a light-cone gauge approach,
where the two light-cone directions (whose oscillators do not create physical states) are
the Cartan direction of $\mathfrak{sl}(2,\mathbb{R})$, as well as the circle direction associated to the ${\rm S}^1$.

\subsection{The spacetime BPS spectrum}

It was shown in \cite{Elitzur:1998mm}, building on \cite{Giveon:1998ns},  that the spacetime theory has a large 
${\cal N}=4$ superconformal symmetry. The large ${\cal N}=4$ superconformal algebra
is generated by the Virasoro algebra --- the asymptotic symmetry algebra arising from AdS$_3$ ---
as well as two affine $\mathfrak{su}(2)$ Kac-Moody algebras that arise from the two 
${\rm S}^3$ factors.  Furthermore there is a $\mathfrak{u}(1)$ algebra corresponding to the
${\rm S}^1$.\footnote{For the convenience of the reader we have reviewed the structure of the algebra and 
its BPS representations in Appendix~\ref{app:Agamma}.} In the unflowed sector, the levels of the two $\mathfrak{su}(2)$ algebras
of the spactime ${\cal N}=4$ superconformal algebra can be identified with the levels of the two $\mathfrak{su}(2)$ algebras
on the world-sheet, while the central charge is equal to $c=6k$, with $k$ the level of the $\mathfrak{sl}(2,\mathbb{R})$
WZW model \cite{Giveon:1998ns,Elitzur:1998mm}.
The BPS bound of the large ${\cal N}=4$ superconformal algebra has the form 
\cite{Gunaydin:1988re,Petersen:1989zz,Petersen:1989pp}, see also \cite{de Boer:1999rh}
\begin{equation}\label{AgBPS}
h \geq \frac{1}{k^++k^-} \, \Bigl[ k^+ j^- + k^- j^+ + u^2 + (j^+-j^-)^2 \Bigr] \ ,
\end{equation}
where $h$ is the conformal dimension, while $j^\pm$ are the spins with respect to the two 
$\mathfrak{su}(2)$ algebras. Furthermore, $u$ is the $\mathfrak{u}(1)$ charge. In the following
we shall only consider the neutral sector $u=0$ --- we have checked that there are no BPS states
for $u\neq 0$. 

In terms of world-sheet parameters, we can identify $h$ with the eigenvalue $j$ of ${\cal J}_0^3$, 
while $j^\pm$ are the spins of ${\cal K}_0^{\pm,a}$. We want to ask which physical states
of the world-sheet theory saturate the BPS bound (\ref{AgBPS}). In order to identify the potential 
BPS states we will fix $j^\pm$, and look for the physical state with the lowest value of $j$. The eigenvalues
$j^\pm$ differ from those of the ground states $j_0^\pm$ by the charges that are carried by the oscillators,
and we define 
\be
j^\pm = j_0^\pm - \Delta^{\pm} \ . 
\ee
If $\Delta^\pm=0$, then we can use all the oscillators to lower $j$, leading to 
\be\label{jmin}
j = j_0 - N - \frac{1}{2} \ . 
\ee
[Note that $N$ has to be half-integer (because of the GSO projection); furthermore, for 
$N>\frac{1}{2}$ we can use the fermionic mode $\psi^{-}_{-1/2}$ once, but then it becomes more
efficient to use the bosonic $J^{-}_{-1}$ modes.] 
On the other hand, if $\Delta^+\neq 0$, we need $|\Delta^+|-\tfrac{1}{2}$ oscillators to obtain the correct
$j^+$ spin and similarly for $j^-$, and then only the remaining oscillators can be used to 
reduce the eigenvalue of $j$. Thus it seems plausible --- and it is not hard to show rigorously, although
the argument is a bit tedious --- that the BPS states can only occur for $\Delta^+=\Delta^-=0$ and 
$N=\tfrac{1}{2}$. (Note that this is also what one would have guessed on general grounds since $N=\tfrac{1}{2}$ characterises 
the `supergravity' states.) Then $j^\pm = j_0^\pm$, and determining $j_0$ from the mass-shell condition and 
plugging it into (\ref{jmin}) leads to 
\begin{align}
j & = - \frac{1}{2} + \sqrt{ \frac{1}{4} +  \frac{k\, j^+ (j^+ + 1)}{k^+} + \frac{k\, j^- (j^- + 1)}{k^-} } \nonumber \\
& = - \frac{1}{2} + \sqrt{ \frac{1}{4} +  \frac{k^-\, j^+ (j^+ + 1)}{k^++k^-} + \frac{k^+\, j^- (j^- + 1)}{k^++k^-} }  \ ,\label{main}
\end{align}
where we have used that $k$ is given by, see eq.~(\ref{krel})
\be
k = \frac{k^+ k^-}{k^+ + k^-} \ . 
\ee
One can convince oneself that (\ref{main}) satisfies the BPS bound (\ref{A28}) 
\be\label{BPSA}
j \geq \frac{1}{k^++k^-} \, \Bigl[ k^- j^+ + k^+ j^- +  (j^+-j^-)^2 \Bigr] 
\ee
provided that $j_0\leq \frac{k+1}{2}$; this latter condition is a consequence of the no-ghost theorem 
\cite{Evans:1998qu,Maldacena:2000hw} and guarantees unitarity. More specifically, after squaring 
$(j+\frac{1}{2})$ from (\ref{main}) and comparing to (\ref{BPSA}), the BPS bound becomes
\be
(j^+-j^-)^2 \, \Bigl[ k^+k^--k^+-k^--(j^+-j^-)^2-2(k^-j^++k^+j^-) \Bigr] \geq 0 \ . 
\ee
The first factor is clearly non-negative, while the square bracket defines an ellipse in the $(j^+,j^-)$-plane. One can see 
that this lies outside the variety --- this is just the mass-shell condition for the maximal choice of $j_0=\frac{k+1}{2}$ --- 
\begin{align}
-\frac{k^2-1}{4k}+\frac{j^+(j^++1)}{k^+}+\frac{j^-(j^-+1)}{k^-}=0 \ , 
\end{align}
but they touch at the point 
\begin{align}
j^+=j^-=\frac{k-1}{2}\ .
\end{align}
Thus the BPS bound is always satisfied, but it can (and is) only saturated for the case
\be
j^+ = j^- \ . 
\ee
This is the main result of our world-sheet analysis.

We have also performed a similar analysis for the spectrally flowed sectors in the NS sector. They give rise to further BPS states, as explained in detail in \cite{Eberhardt:2017pty}. However, since supergravity corresponds to the regime $k^\pm \to \infty$, these additional states are not important for the present analysis. Thus from now on we will only talk about the unflowed sector of string theory.

As regards the situation in the R sector, the analysis is similar to the above, and the 
only BPS states exist for $j^+=j^-$. 
The corresponding values of $j^+=j^-$ are shifted by $\frac{1}{2}$ relative to the NS analysis. These 
BPS states are therefore the states associated to the $j+\frac{1}{2}$ term in (\ref{largeN4BPS}), and they are required in order 
to get the complete multiplet of the large ${\cal N}=4$ superconformal algebra. 
Thus our analysis predicts that the entire BPS spectrum of string theory on 
AdS$_3 \times {\rm S}^3 \times {\rm S}^3 \times {\rm S}^1$ consists of the representations 
\be\label{BPSstring}
\hbox{BPS spectrum of string theory:} \quad \bigoplus_{j=0}^{ \frac{k-1}{2}} \, [j,j,u=0]_S \otimes \overline{[j,j,u=0]}_S \ . 
\ee

\subsection{Supergravity interpretation}\label{sec:sugraint}

Given the general relation between string theory and supergravity, one should expect
that (\ref{main}) also has a direct supergravity interpretation. Let us consider a scalar field on AdS$_3$ that 
arises from a massless scalar field in 10 dimensions
upon KK reduction on ${\rm S}^3 \times {\rm S}^3 \times {\rm S}^1$. If we take the KK momentum along
the ${\rm S}^1$ to be trivial, then its mass (in AdS$_3$ units, i.e, relative to the size $k$), is 
\be
\frac{m^2}{ k} =  \frac{j^+ (j^+ + 1)}{k^+} + \frac{j^- (j^- + 1)}{k^-} \ ,
\ee
where the two terms on the right-hand-side are the eigenvalues of the Laplacian on the sphere for the 
spherical harmonic labelled by $(j^+,j^-)$. Again these eigenvalues are evaluated in appropriate units,
i.e., relative to the sizes $k^\pm$ of the corresponding ${\rm S}^3$'s. 
We can convert this expression into the conformal dimension of the dual CFT, using the general relation between the 
mass of a scalar field and the left-moving conformal dimension $\Delta = h + \bar{h}$ (with $h=\bar{h}$),
\be
m^2 = h(h - 1) \ ,
\ee
leading to 
\be\label{sugra1}
h= \frac{1}{2}  +  \sqrt{ \frac{1}{4} + m^2} = 
\frac{1}{2} + \sqrt{ \frac{1}{4} +  \frac{k\, j^+ (j^+ + 1)}{k^+} + \frac{k\, j^- (j^- + 1)}{k^-} } \ .
\ee
This differs by a shift of $1$ from (\ref{main}) --- as we shall see in the next section, the analysis is a bit
more subtle, and there is in fact one scalar component for which (\ref{main}) is precisely reproduced --- 
but the dependence on the spins is exactly as in (\ref{main}). For the supergravity analysis the relevant
symmetry algebra is $D(2,1|\alpha)$, whose BPS bound takes the form (\ref{A30}) (see Appendix~\ref{app:D21}) 
\be\label{BPSD} 
h \geq \frac{1}{k^++k^-} \, \Bigl[ k^- j^+ + k^+ j^-  \Bigr] \ .
\ee
Provided that the actual supergravity analysis leads to (\ref{sugra1}) (or rather to $h-1$), it follows by similar arguments
as above that it can only be saturated if $j^+=j^-$. Indeed the supergravity bound (\ref{BPSD}) is weaker than the stringy bound
(\ref{BPSA}), and hence at most those states that saturate the stringy bound can saturate the supergravity bound. Happily,
the only states that saturate the stringy bound occur for $j^+=j^-$ where the two bounds coincide.
\smallskip

If this somewhat sketchy line of reasoning is correct it would suggests that the BPS spectrum of supergravity on 
AdS$_3 \times {\rm S}^3 \times {\rm S}^3 \times {\rm S}^1$ consists only of representations of 
$D(2,1|\alpha)$ with $j^+=j^-$.  This is contrary to what was claimed or assumed in the literature before,
see in particular \cite{de Boer:1999rh}. We shall therefore, in the next section, perform a careful and detailed 
supergravity computation to confirm this claim explicitly.

\section{Supergravity approach}\label{sec:sugra}

Let us start with $10$-dimensional IIB supergravity. The 
bosonic part of the action is, in the string frame, given by
\begin{align}
S_\mathrm{IIB}&=S_\mathrm{NS}+S_\mathrm{R}+S_\mathrm{CS} \nonumber\\
S_\mathrm{NS}&=\frac{1}{2\kappa^2} \int \mathrm{d}^{10}x \sqrt{-g} e^{-2\Phi}\left(R+4 \partial_M \Phi \partial^M \Phi -\frac{1}{2} \abs{H}^2\right)\ ,\nonumber \\
S_\mathrm{R}&=-\frac{1}{4\kappa^2} \int \mathrm{d}^{10}x \sqrt{-g} \left(\abs{F_1}^2+|\tilde{F}_3|^2+\frac{1}{2}|\tilde{F}_5|^2\right)\ , \nonumber\\
S_\mathrm{CS}&=-\frac{1}{4\kappa^2} \int C_4 \wedge H \wedge F_3\ , \label{IIB_action}
\end{align}
where
\begin{align}
\tilde{F}_3&=F_3-C_0 \wedge H \ , \\
\tilde{F}_5&=F_5-\frac{1}{2} C_2 \wedge H +\frac{1}{2} B_2 \wedge F_3 \ .
\end{align}
Here, $\Phi$ is the 10d dilaton field, $g$ is the 10d metric with $R$ its associated curvature scalar, 
and $B$ is the Kalb-Ramond field. $C_0$, $C_2$ and $C_4$ are the R-R-fields, whose fields strengths are defined as
\be 
F_1= \mathrm{d} C_0\ , \quad H = \mathrm{d} B\ , \quad F_3 = \mathrm{d} C_2 \ , \quad F_5 = \mathrm{d} C_4 \ .
\ee
Finally, we have to impose self-duality on $\tilde{F}_5$, 
\be 
\star \tilde{F}_5 =\tilde{F}_5 \ .
\ee 
We will look at solutions on $\mathrm{AdS}_3 \times \mathrm{S}^3_+ \times \mathrm{S}^3_- \times \mathrm{S}^1$ where we have 
pure NS-flux on the AdS$_3$ factor and through the two  $\mathrm{S}^3_\pm$.

It is consistent to set all R-R fields (as well as all the fermions) to zero. Then we are only left with the 
NS-NS fields, i.e.,  the metric, the dilaton and $H$, whose action is 
\begin{align}
\tilde{S}_\mathrm{IIB}=\frac{1}{4\kappa^2} \int \mathrm{d}^{10}x\ \sqrt{-g} \, e^{-2\Phi}
\left(R+4\partial_M \Phi \partial^M \Phi-\frac{1}{2}\abs{H}^2\right)\ .
\end{align}

\subsection{The 9d vacuum solution}\label{sec:3.1}
We now compactify the theory on a circle to get an effective 9d theory. In the process, we have to dimensionally
reduce the fields, see, e.g., \cite{ortin2004gravity}.
The 10d metric $g$ gives rise to a 9d-metric, which we again call $g$, a vector $A_\mu=g_{\mu,10}$ and a scalar 
$k=g_{10,10}$. The Kalb-Ramond field $B$ leads to a vector $\hat{A}_\mu=B_{\mu,10}$ and a 9d two-form $B$. We call the field 
strengths associated to $A$ and $\hat{A}$, $F$ and $\hat{F}$, respectively. The resulting action is then --- this is eq.~(21.25) 
of \cite{ortin2004gravity}, after changing the signature, rescaling the fields and setting $\Psi=\log k$, 
\begin{align}
S_{\mathrm{9d}}  =  \frac{1}{4\kappa^2}\int \mathrm{d}^9 x \sqrt{-g}\, e^{-2\Phi} \,  
\Bigl[ & R+4 \partial_M \Phi \partial^M\Phi -\frac{1}{2}\abs{H}^2-\partial_M \Psi \partial^M \Psi\nonumber \\
& -\frac{1}{2}\abs{F}^2-\frac{1}{2}|\hat{F}|^2\Bigr]\ .
\end{align}
The result is invariant under T-duality, which interchanges $A$ and $\hat{A}$.
The field equations for $\Psi$ and the one-forms read
\be 
\Delta A=\Delta \hat{A}=0\ , \quad \Delta \Psi=0\ .
\ee
It is hence consistent to set them all to zero. 
The $\Phi$-equation of motion is
\be 
R-\frac{1}{2} \abs{H}^2=4 \partial_M \Phi \partial^M \Phi +4 \Delta \Phi \ .
\ee
We set $\Phi=0$ (or $\Phi$ at least constant), which imposes the condition
\be 
R=\frac{1}{2} \abs{H}^2\ . \label{phi_constraint}
\ee
The $g$-equation of motion is the usual Einstein equation,
\be 
R_{MN}-\frac{1}{4} H_{MPR} \tensor{H}{_N^{PR}}=0\ .
\ee
where we have used (\ref{phi_constraint}).
Taking the trace gives
\be 
R=\frac{1}{4}\abs{H}^2\ , \label{trace_Einstein_equation}
\ee
and hence from \eqref{phi_constraint} and \eqref{trace_Einstein_equation} we must have $R= \abs{H}^2=0$. 
The $H$ equation of motion finally reads
\be 
\mathrm{d}\star H=0\ ,
\ee
i.e. $H$ is closed and coclosed and hence harmonic. 

In order to find the vacuum solution corresponding to AdS$_3\times {\rm S}^3_+ \times {\rm S}^3_-$, let us denote 
 their radii by $r$ (for $\mathrm{AdS}_3$), and $r_{\pm}$ (for $\mathrm{S}_{\pm}^3$). All factors are maximally symmetric, 
 and thus the Riemann tensor takes the form 
\be 
R_{MNPR} \propto \left(g_{MP}g_{NR}-g_{MR}g_{NP}\right)\ ,
\ee
if all indices are in one factor. We thus have
\begin{align}
R_{MNPR}&=-r^{-2} \left(g^{\mathrm{AdS}_3}_{MP}g^{\mathrm{AdS}_3}_{NR}-g^{\mathrm{AdS}_3}_{MR}g^{\mathrm{AdS}_3}_{NP}\right)
+r_+^{-2} \left(g^{\mathrm{S}^3_+}_{MP}g^{\mathrm{S}^3_+}_{NR}-g^{\mathrm{S}^3_+}_{MR}g^{\mathrm{S}^3_+}_{NP}\right)\nonumber\\
&\qquad\qquad+r_-^{-2} \left(g^{\mathrm{S}^3_-}_{MP}g^{\mathrm{S}^3_-}_{NR}-g^{\mathrm{S}^3_-}_{MR}g^{\mathrm{S}^3_-}_{NP}\right)\ ,
\end{align}
where we view $g^{\mathrm{AdS}_3}$ etc.\ as 10d fields, i.e., $g^{\mathrm{AdS}_3}_{MN}=0$ except when $0 \le M,N \le 2$, etc. 
Contracting indices gives
\begin{align}
R_{MN}&=-2r^{-2} g_{MN}^{\mathrm{AdS}_3}+2r_+^{-2} g_{MN}^{\mathrm{S}^3_+}+2r_-^{-2} g_{MN}^{\mathrm{S}^3_-}\ , \\
R&=-6r^{-2}+6 r_+^{-2}+6 r_-^{-2} \ .
\end{align}
Following \cite{Gukov:2004ym}, we now make the ansatz for $H$ 
\be 
H=\lambda_0 \omega^{\mathrm{AdS}_3}+\lambda_+ \omega^{\mathrm{S}^3_+}+\lambda_- \omega^{\mathrm{S}^3_-}\ ,
\ee
where $\omega^{\mathrm{AdS}_3}$ are the volume forms on $\mathrm{AdS}_3$, etc. This form is trivially closed. The volume forms are harmonic and hence also coclosed, so $H$ obeys the equation of motion. The normalisations are chosen as
\be 
\int_{\mathrm{S}^3_\pm} \omega_\pm = 2\pi^2 r_\pm^3\ ,
\ee
and similarly for $\mathrm{AdS}_3$, which can be obtained by analytic continuation. We 
have $|\omega^{\mathrm{AdS}_3}|^2=-1$ and $|\omega^{\mathrm{S}^3_\pm}|^2=1$, i.e. in total:
\be 
\abs{H}^2=-\lambda_0^2+\lambda_+^2+\lambda_-^2 \ .
\ee
The Einstein equations exhibit a connection between the $\lambda$'s and the radii, namely
\begin{equation} 
r^{-2}=\frac{1}{4} \lambda_0^2 \ , \qquad \qquad  r_\pm^{-2} = \frac{1}{4}\lambda_\pm^2  \ ,
\end{equation}
and furthermore the vanishing of $R$ implies
\be 
-\lambda_0^2+\lambda_+^2+\lambda_-^2=4(-r^{-2}+r_+^{-2}+r_-^{-2})=0 \ . \label{vanishing_R}
\ee
The fluxes through $\mathrm{S}^3_\pm$ are given by
\be 
\int_{\mathrm{S}^3_\pm} H=Q_5^\pm=k^\pm=4\pi^2 r_\pm^2\ , \label{radius_level}
\ee
and are always integers. As a consequence, \eqref{vanishing_R} coincides with the string theory criticality condition on the levels,
see eq.~(\ref{krel})
\be 
\frac{1}{k}=\frac{1}{k^+}+\frac{1}{k^-} \ ,\label{level_relation}
\ee
where we defined in analogy $k$ as $k=4\pi^2 r^2$. 

\subsection{Equations of motion for quadratic perturbations}\label{sec:3.2}

We now consider the fluctuations around this background geometry. To second order in the variations the above action becomes
\be 
\delta^2 \tilde{S}_\mathrm{IIB}=\frac{1}{4\kappa^2}\int \mathrm{d}^{9}x\ \sqrt{-g} \, \mathscr{L}\ ,
\ee
where $\mathscr{L}$ is explicitly given as 
\begin{align}
\mathscr{L}&=\frac{1}{2}\nabla_P \tensor{h}{^R_R}\nabla^P \tensor{h}{^M_M}-\frac{1}{2}\nabla_R h_{MP}\nabla^R h^{MP}+\nabla_P h_{MR}\nabla^Rh^{MP}- \nabla_P \tensor{h}{^R_R}\nabla^R \tensor{h}{^P_R}\nonumber\\
&\qquad-8\nabla_P\phi \nabla^{[P}\tensor{h}{^{M]}_M}-2 \nabla_M \psi \nabla^M\psi+8 \nabla_M \phi \nabla^M\phi+\frac{1}{6}H_{MPR}\Xi^{MPR}(4\phi-\tensor{h}{^Q_Q}) \nonumber\\
&\qquad+\tensor{H}{_M^{RQ}}h^{MP}\Xi_{PRQ}-\frac{1}{6}\Xi_{MPR}\Xi^{MPR}-\frac{1}{2}Z_{MP}Z^{MP}-\frac{1}{2}\hat{Z}_{MP}\hat{Z}^{MP}\nonumber\\
&\qquad-\frac{1}{2}\tensor{H}{_M_R^N}\tensor{H}{_P_Q_N}h^{MP}h^{RQ}\ . \label{action_fluctuations_simplified}
\end{align}
Here we wrote $\tensor{\delta g}{_M_N}=h_{MN}$, and we treat $h_{MN}$ as a tensor, i.e., $\delta g^{MN}=-h^{MN}$. We similarly defined
$\delta H_{MNP}=\Xi_{MNP}$, $\delta \Phi=\phi$, $\delta \Psi=\psi$, $\delta F_{MN}=Z_{MN}$, as well as $\delta \hat{F}_{MN}=\hat{Z}_{MN}$. The first terms are kinetic terms, but there are also mass terms. The 
connection is still the Levi-Civita connection of the background metric, and we raise and lower indices with the background metric. We 
used that $\mathrm{AdS}_3 \times \mathrm{S}^3_+ \times \mathrm{S}^3_-$ has vanishing Ricci scalar and vanishing $\abs{H}^2$;
we have also already inserted the other background values of the fields and used the Einstein equation of the background solution. 
Note that there are no interference terms between $Z$, $\hat{Z}$, $\psi$, and the other fields, and hence we  can
treat the fluctuations of $F$, $\hat{F}$ and $\Psi$ separately from the rest. 
The resulting equations of motion for the remaining quadratic fluctuations are
\begin{align}
0&=-\Delta \phi -\frac{1}{4}\nabla_M \nabla_P h^{MP}+\frac{1}{4}\Delta \tensor{h}{^M_M}+\frac{1}{24}H^{MPQ}\Xi_{MPQ} \label{phi_eom}\ , \\
0&=-\Delta h_{MP}-4 g_{MP}\Delta \phi -g_{MP}\nabla_R\nabla_Q h^{QR}+g_{MP}\Delta \tensor{h}{^Q_Q}
-\nabla_M \nabla_P \tensor{h}{^Q_Q}\nonumber\\
&\qquad +2\nabla_Q \nabla_{(M}\tensor{h}{_{P)}^Q}
+4 \nabla_M\nabla_P\phi +\frac{1}{6}g_{MP}H^{QRN}\Xi_{QRN}-\tensor{H}{_M^Q^R}\Xi_{PQR}
\nonumber \\
&\qquad  +\tensor{H}{_{MQ}^N}H_{PRN}h^{QR}\ , \label{g_eom} \\[4pt]
0&=-\Delta X_{MP}+2\nabla_Q \nabla_{[P}\tensor{X}{_{M]}^Q}+2\phi \nabla_Q \tensor{H}{_M_P^Q}
+2H_{MPQ}\nabla^Q\phi +h^{QR}\nabla_R H_{MPQ}\nonumber\\
&\qquad-\frac{1}{2}\tensor{h}{^Q_Q}\nabla_R\tensor{H}{_M_P^R}-\tensor{h}{^Q_M}\nabla_R\tensor{H}{_P_Q^R}+H_{MPQ}\nabla_R h^{QR}-H_{PQR}\nabla^R \tensor{h}{^Q_M}\nonumber\\
&\qquad+H_{MQR}\nabla^R\tensor{h}{^Q_P}-\frac{1}{2}\nabla^R \tensor{h}{^Q_Q}\ ,\label{X_eom}
\end{align}
where $\delta B_{MN}=X_{MN}$. 

\subsection{Expanding the Fields}\label{sec:3.3}
Following \cite{Deger:1998nm}, we parametrise the metric fluctuations as 
\begin{align} 
\delta g_{\mu\nu}&=H_{\mu\nu}+g_{\mu\nu} M\ , \quad g^{\mu\nu}H_{\mu\nu}=0\ , \\
\delta g_{\mu a}&=R_{\mu a} \ , \\
\delta g_{\mu i}&=S_{\mu i} \ , \\
\delta g_{a i}&=T_{a i} \ , \\
\delta g_{ab}&=K_{ab}+g_{ab} N \ , \quad g^{ab}L_{ab}=0\ , \\
\delta g_{i j }&=L_{ij}+g_{ij} P \ , \quad g^{ij}T_{ij}=0\ .
\end{align}
Here and from now on, greek indices refer to $\mathrm{AdS}_3$, latin indices from the beginning of the alphabet 
$a,b, \ldots$ to $\mathrm{S}^3_+$,  while the latin indices from the middle of the alphabet $i,j,\ldots$ refer to $\mathrm{S}^3_-$. 
We will use capital latin letters to indicate 9d-indices. 
For the antisymmetric tensor field, we have $H_{MNP}=3\partial_{[M} B_{NP]}$, and we parametrise the fluctuations as 
\begin{align} 
X_{\mu\nu}&=\epsilon_{\mu\nu\rho} U^\rho\ , \\
X_{a b}&=\epsilon_{abc} V^c \ , \\
X_{ij}&=\epsilon_{ijk} W^k\ , \\
X_{\mu a}&=C_{\mu a} \ ,  \\
X_{\mu i}&= D_{\mu i} \ , \\
X_{a i}&= E_{a i} \ .
\end{align}
We expand the fluctuations in harmonic functions on the two $\mathrm{S}^3$'s as 
\begin{align}
\phi&=\sum\nolimits_{\ell^+,\ell^-} \phi^{(\ell^+\, 0) (\ell^-\, 0)} \, Y_+^{(\ell^+\, 0)}Y_-^{(\ell^-\, 0)}\ , \label{harmonic_expansion_phi}\\
V_{a}&=\sum\nolimits_{\ell^+,\ell^-} V^{(\ell^+\, \pm 1) (\ell^-\, 0)} \, Y_{+,a}^{(\ell^+\, \pm 1)}Y_-^{(\ell^-\, 0)}
+V^{(\ell^+\, 0)(\ell^-\, 0)} \, \partial_a Y_+^{(\ell^+\, 0)}Y_-^{(\ell^-\, 0)}\ , \\
K_{ab}&=\sum\nolimits_{\ell^+,\ell^-} K^{(\ell^+\, \pm 2)(\ell^-\, 0)} \, Y_{+,ab}^{(\ell^+\, \pm 2)}Y_-^{(\ell^-\, 0)}
+K^{(\ell^+\, \pm 1)(\ell^-\, 0)} \,  \nabla_{\{a} Y_{+,b\}}^{(\ell^+\, \pm 1)}Y_-^{(\ell^-\, 0)}\nonumber\\
&\qquad\qquad+K^{(\ell^+\, 0)(\ell^-\, 0)}\, \nabla_{\{a}\nabla_{b\}} Y_+^{(\ell^+\, 0)}Y_-^{(\ell^-\, 0)}\ ,\label{harmonic_expansion_K}
\end{align}
where we have chosen, as representative examples, fields of spin $0$, $1$ and $2$ on $\mathrm{S}^3_+$, respectively; the 
complete list for all the fields is given in Appendix~\ref{app:harmonics}. Here, $\{ab\}$ resp.\ $\{ij\}$ denotes the traceless symmetric part, 
and $Y^{(\ell_1\, \ell_2)}_{\pm,(s)}$ is the  
eigenfunction of the Laplacian on $\mathrm{S}^3_\pm$, which transforms under 
$\mathrm{SO}(4) \cong \mathrm{SU}(2) \times \mathrm{SU}(2)$ in the representation 
\be 
\frac{1}{2}(\ell_1+\ell_2)\ , \quad \frac{1}{2}(\ell_1-\ell_2)\ .
\ee
Here $\ell_i\in\mathbb{N}_0$, and the relation to the previously used $j$ variables is 
\be\label{trans}
j=\frac{1}{2}(\ell_1+\ell_2)\ , \quad \bar{j}=\frac{1}{2}(\ell_1-\ell_2)\ .
\ee
The space of $1$-forms on $\mathrm{S}^3_+$ is spanned by
\be 
Y_{+,a}^{(\ell\, \pm 1)}\ , \quad \partial_a Y_+^{(\ell\, 0)}
\ee
and the space of traceless symmetric 2-tensors is spanned by
\be 
Y_{+,ab}^{(\ell\,\pm 2)}\ , \quad \nabla_{\{a}Y_{+,b\}}^{(\ell\,\pm 1)}\ , \quad \nabla_{\{a}\nabla_{b\}} Y_+^{(\ell\, 0)}\ .
\ee
As in \cite{Deger:1998nm}, we choose the `Lorentz gauge' 
\be 
\nabla^a h_{a\mu}=0\ , \quad \nabla^a h_{\{ab\}}=0\ , \quad \nabla^a h_{ai}=0\ , \quad \nabla^a X_{a M}=0\ . \label{gauge_conditions}
\ee
This gauge removes $9$ degrees of freedom of the metric and $8$ degrees of freedom of $X$, which are the right numbers. Hence locally, 
this gauge is admissible, and one can also confirm this more carefully. We should note that this gauge choice breaks the manifest
symmetry between the two spheres (since we impose the divergence condition only with respect to the $a$-coordinates on 
${\rm S}^3_+$). However, as we shall see later, the resulting spectrum will be symmetrical with respect to exchanging the two spheres.
We should note that eq.~(\ref{gauge_conditions}) implies in particular that 
\begin{align}
\nabla^a R_{\mu a}&=0\ , \quad \nabla^a K_{ab}=0\ , \quad \nabla^a  T_{a i}=0\ , \\
\nabla^a C_{\mu a}&=0\ , \quad \nabla_{[a} V_{b]}=0\ , \quad \nabla^a E_{a i}=0\ . 
\end{align}
Additionally, there is a gauge freedom in the decoupled subsystem which will be fixed in the next section.
\subsection{The scalars from the decoupled subsystem}\label{sec:decoupled}
As was mentioned before, we can analyse the fluctuations of $\Psi$, $F$ and $\hat{F}$ separately from the rest;
since this part of the analysis is simpler, let us first explain that. The equation of motion of $\Psi$ is simply
\be 
(\Delta_0+\Delta_++\Delta_-)\Psi=0\ . \label{eom_psi}
\ee
Expanding $\Psi$ in terms of harmonics, 
\be
\Psi=\sum\nolimits_{\ell^+,\ell^-} \Psi^{(\ell^+\, 0) (\ell^-\, 0)} \, Y_+^{(\ell^+\, 0)}Y_-^{(\ell^-\, 0)}\ , 
\ee
and using \eqref{harmonic_function_scalars}, we get
\be
(\Delta_0 - m^2_{\ell^+,\ell^-} ) \, \Psi^{(\ell^+\, 0) (\ell^-\, 0)} = 0 \ , 
\ee
where 
\be \label{mass1}
m^2_{\ell^+,\ell^-} \equiv r_+^{-2}\ell^+(\ell^++2)+r_-^{-2}\ell^-(\ell^-+2)\ ,
\ee 
is the $\mathrm{AdS}_3$ mass squared.
The treatment of $Z_{MN}=2\partial_{[M}\Upsilon_{N]}$ is a bit more complicated. Let
us denote the perturbation of $A_M$ as $\Upsilon_M$, and split $\Upsilon_M$ as 
\be 
\Upsilon_\mu=\Pi_\mu\ , \quad \Upsilon_a =\Sigma_a\ , \quad \Upsilon_i=\Omega_i\ .
\ee
To fix the gauge, we require the equivalent of the last equation of eq.~\eqref{gauge_conditions}
\be 
\nabla^a \Sigma_a=0\ .
\ee
The equations of motion, split into the different components, read
\begin{align}
0&=(\Delta_0+\Delta_++\Delta_-+2\ell^{-2})\Pi_\mu-\nabla_\mu\nabla^\nu \Pi_\nu -\nabla_\mu \nabla^i \Omega_i\ , \label{eom_pi}\\
0&=(\Delta_0+\Delta_++\Delta_--2r_+^{-2})\Sigma_a-\nabla_a\nabla^\nu \Pi_\nu -\nabla_a \nabla^i \Omega_i\ , \label{eom_sigma}\\
0&=(\Delta_0+\Delta_++\Delta_--2r_-^{-2})\Omega_i-\nabla_i\nabla^\nu \Pi_\nu -\nabla_i \nabla^j \Omega_j\ . \label{eom_omega}
\end{align}
We shall in the following always focus on scalar quantities, i.e., on modes that are scalar with respect to 
$\mathrm{AdS}_3$ as well as the two $\mathrm{S}^3_\pm$. In the present context, there are two of those, 
namely $\nabla^\mu \Pi_\mu$ and $\nabla^i\Omega_i$. Thus, we have an overconstrained system, since the scalar parts of 
\eqref{eom_pi} -- \eqref{eom_omega} constitute three equations for these two fields. Let us extract the scalar parts of these three 
equations by applying $\nabla^\mu$ to the first equation, $\nabla^a$ to the second one and $\nabla^i$ to the third one; the result is
\begin{align}
0&=(\Delta_++\Delta_-)\nabla^\mu \Pi_\mu-\Delta_0 \nabla^i\Omega_i\ , \\
0&=\Delta_+(\nabla^\nu \Pi_\nu+\nabla^i\Omega_i)\ , \\
0&=(\Delta_0+\Delta_+)\nabla^\mu \Omega_i-\Delta_-\nabla^\mu \Pi_\mu\ .
\end{align}
For $\ell^+>0$, the second equation implies that there is actually only one scalar field, which we may take to 
be $\nabla^\mu\Pi_\mu$;  for $\ell^+=0$, the situation is more complicated and requires fixing the residual gauge, as
explained in Appendix~\ref{app:special}. Assuming $\ell^+>0$, the first and third equation are equivalent to
\be 
0=(\Delta_0+\Delta_++\Delta_-)\nabla^\mu\Pi_\mu\ .
\ee
(Later, we will also have a similar situation for the dilaton, the metric and the Kalb-Ramond field, i.e., 
the overconstrained system of scalar equations will imply some algebraic relationships among the fields, and once we have found these, 
there are some linear dependencies among the remaining equations.) We hence get again one set of scalar fields with mass
\be \label{mass2}
m^2 = m^2_{\ell^+,\ell^-}  \equiv r_+^{-2}\ell^+(\ell^++2)+r_-^{-2}\ell^-(\ell^-+2)\ .
\ee
Thus in total we find three sets of scalar fields, one from $\Psi$, one from $A_M$ and one from $\hat{A}_M$, all having equal mass. 

\subsection{The scalars from the remaining fields}\label{sec:remain}

The analysis of the remaining fluctuations is more complicated, and the details are spelled out in Appendix~\ref{app:sugra}
(and have been largely performed with the help of {\tt Mathematica}, see also the ancillary workbook of the {\tt arXiv} submission).
It follows from eqs.~({\ref{B57}) -- (\ref{B63}) that all the remaining fields mix, but one can 
diagonalize the corresponding matrix. The eigenvectors are complicated, but the eigenvalues --- they correspond to the masses of the particles, since the 
above system of equation is simply the Klein-Gordon equation of seven coupled scalar particles on $\mathrm{AdS}_3$
--- are quite simple. Five of the eigenvalues of the matrix are directly of the above form 
\begin{align}
\lambda_1&=m^2_{\ell^+,\ell^--2}\ , \\
\lambda_2&=m^2_{\ell^+-2,\ell^-}\ , \\
\lambda_3&=m^2_{\ell^+,\ell^-}\ , \\
\lambda_4&=m^2_{\ell^+,\ell^-+2}\ , \\
\lambda_5&=m^2_{\ell^++2,\ell^-}\ ,
\end{align}
where $m^2_{\ell^+,\ell^-}$ is defined in (\ref{mass1}), see also (\ref{mass2}).
The remaining two eigenvalues are more complicated and take the form 
\begin{align}
\lambda_{6/7}&=\left(2r^{-1}\pm \sqrt{r^{-2}+m^2_{\ell^+,\ell^-}}\right)^2-r^{-2} \ . \label{lambda67}
\end{align}
Hence their associated conformal dimensions, which are obtained via
$r^2 \lambda^2 = h (h-1)$, are
\begin{align}\label{lam67}
h_{6/7}=\frac{1}{2}\left(1+\sqrt{1+r^2 \lambda_{6/7}}\right) =h_{\ell^+,\ell^-}\pm 1\ ,
\end{align}
where $h_{\ell^+,\ell^-}$ is the conformal dimension associated to $m^2_{\ell^+,\ell^-}$,
\begin{align}
h_{\ell^+,\ell^-} & =  \frac{1}{2} + \frac{1}{2} \, \sqrt{1+r^2 \left(\frac{\ell^+(\ell^++2)}{r_+^2}+\frac{\ell^-(\ell^-+2)}{r_-^2}\right)}\nonumber \\[4pt]
& =\frac{1}{2}+\sqrt{\frac{1}{4}+k \left(\frac{j^+(j^++1)}{k^+}+\frac{j^-(j^-+1)}{k^-}\right)} \equiv  h_{j^+,j^-}  \ .
\end{align}
In the final step we have used eq.~(\ref{trans}) to convert the integer valued labels $\ell^\pm$ into the spin labels $j^\pm$, and 
eq.~(\ref{radius_level}) to change from the radii to the levels we used earlier.

\subsection{The full scalar spectrum}\label{sec:full}

Taking these results together, we thus conclude that the supergravity spectrum contains the ten scalar fields
\begin{align}
& (h_{j^+ \pm 1,j^-};j^+,j^-)\ , \quad  (h_{j^+,j^- \pm 1};j^+,j^-)\ ,  
& 4 \times (h_{j^+,j^-}; j^+,j^-)\ , \quad (h_{j^+,j^-} \pm 1;j^+,j^-)\ .
\label{scalar_spectrum}
\end{align}
Note that we have suppressed $\bar{h}$ and $\bar{j}^\pm$, since they agree with $h$ and $j^\pm$, respectively, as we
are only considering scalars, see the comment after eq.~(\ref{eom_omega}). 
We have furthermore assumed that $j^\pm\geq 1$; for small values of $j^\pm$ (or rather $\ell^\pm$) the above
analysis has to be adjusted, and this is explained in detail in Appendix~\ref{app:special}. (In particular, it follows
that if $j^\pm =\frac{1}{2}$, the fields with spin $j^\pm -1$ are absent; in addition, there are further restrictions if either
(or both) spins vanish, $j^\pm=0$.) The above 
spectrum accounts correctly for the scalar modes of the NS-NS fields; there are also scalar fields arising from the 
R-R fields.

We note that $h_{j^+,j^-}$ agrees precisely with our naive prediction from above, see eq.~(\ref{sugra1}). As was explained there,
this differs by $1$ from (\ref{main}); thus the only state that can directly satisfy the BPS bound is the one corresponding to the
eigenvalue $h_{j^+,j^-}- 1$, and it actually only saturates the bound if $j^+=j^-$. 

We can now compare this to the supergravity spectrum spelled out in \cite{de Boer:1999rh}, where it was claimed that 
it takes the form
\smallskip

\noindent Bosonic: 
\be\label{bosd21}
\bigoplus_{j^+ \geq 0,\ j^- \geq 0}  \Bigl( [j^+, j^-; j^+, j^-]_s \ \ \oplus \   [j^++\tfrac{1}{2}, j^-+\tfrac{1}{2}; j^++\tfrac{1}{2}, j^-+\tfrac{1}{2}]_s \Bigr) \ ,
\ee
except that the multiplet $[0,0;0,0]_s$ does not appear in the sum.
\smallskip

\noindent Fermionic: 
\be\label{fermd21}
\bigoplus_{j^+ \geq 0,\  j^- \geq 0}\Bigl( [j^+, j^-; j^+ + \tfrac{1}{2}, j^- +\tfrac{1}{2}]_s
\ \oplus \ [j^+ +\tfrac{1}{2}, j^- +\tfrac{1}{2}; j^+, j^-]_s \Bigr) \ .
\ee
\smallskip

\noindent 
In either case  $[j_1^+,j_1^-;j_2^+,j_2^-]_s \equiv [j_1^+,j_1^-]_s \otimes [j_2^+,j_2^-]_s$, where the first and second 
factor corresponds to the left- and right-movers, respectively (and our conventions for the $D(2,1|\alpha)$
representations are described in Appendix~\ref{app:D21reps}). Since we have only analysed the
scalar fields from the NS-NS sector, we cannot see {\em all} states from our above analysis; however, each multiplet (with the 
exception of the supergravity multiplet $[\tfrac{1}{2},\tfrac{1}{2};0,0]_s$ and $[0,0;\tfrac{1}{2},\tfrac{1}{2}]_s$) contains at least one NS-NS scalar field, and thus we can deduce the conformal dimensions of the 
relevant multiplets. In particular, in the first sum in (\ref{bosd21}) the ground states are NS-NS scalars, 
and they contain the four modes 
\be\label{BPS1}
\begin{array}{rll}
[j^+,j^-;j^+,j^-]_s : \hspace*{0.5cm}& (h_{j^+,j^-}-1;j^+,j^-)  \hspace*{0.3cm}  & (h_{j^+,j^-};j^+,j^-) \\
& (h_{j^+,j^-};j^+-1,j^-) \qquad & (h_{j^+,j^-};j^+,j^--1) \ .  
\end{array}
\ee
On the other hand, the ground states of the second sum in (\ref{bosd21}) are R-R states, and their NS-NS contributions
are then 
\be
\begin{array}{rll}
[j^++\tfrac{1}{2},j^-+\tfrac{1}{2};j^++\tfrac{1}{2},j^-+\tfrac{1}{2}]_s : \hspace*{0.5cm}
& (h_{j^+,j^-};j^++1,j^-)  \hspace*{0.3cm}   & (h_{j^+,j^-};j^+,j^-+1) \\
& (h_{j^+,j^-};j^+,j^-) \qquad & (h_{j^+,j^-}+1;j^+,j^-) \ .  
\end{array}
\ee
Similarly, the two fermionic multiplets contain one NS-NS scalar each
\be
\begin{array}{rl}
{}[j^+, j^-; j^+ + \tfrac{1}{2}, j^- +\tfrac{1}{2}]_s : \hspace*{0.5cm} & (h_{j^+,j^-};j^+,j^-) \\
{}[j^++ \tfrac{1}{2} , j^-+ \tfrac{1}{2}; j^+ , j^- ]_s : \hspace*{0.5cm} & (h_{j^+,j^-};j^+,j^-)  \ . 
\end{array}
\ee
In particular, it therefore follows that the first multiplet (\ref{BPS1}) is only  BPS if $j^+=j^-$, since only in this
case does $h_{j+,j^-}-1$ saturate the BPS bound; if $j^+\neq j^-$, it is actually not BPS and combines with the second
term in (\ref{bosd21}) to form a long multiplet, see eq.~(\ref{shortlong}). The analysis in the other sectors works similarly
(as does the various shortenings for small spin), and we therefore conclude that the correct supergravity spectrum takes the form 
\begin{align} \label{multiplet_spectrum}
\bigoplus_{j^+= j^-} &
\left([j^+,j^-]_s\oplus [j^++\tfrac{1}{2},j^-+\tfrac{1}{2}]_s\right)\otimes \left([j^+,j^-]_s\oplus [j^++\tfrac{1}{2},j^-+\tfrac{1}{2}]_s\right) \\ 
&\qquad \oplus \ \bigoplus_{j^+ \ne j^-} [j^+,j^-] \otimes [j^+,j^-]\ ,  \nonumber 
\end{align}
except that, as before, see eq.~(\ref{bosd21}), the vacuum term $[0,0]_s\otimes [0,0]_s$ does not appear.

\section{Discussion and Outlook}\label{sec:concl}

In this paper we have analysed the BPS spectrum of supergravity and string theory on 
AdS$_3\times {\rm S}^3 \times {\rm S}^3 \times {\rm S}^1$ (with pure NS-NS flux). We have found that
both spectra only contain BPS states in representations for which $j^+=j^-$. Furthermore, the BPS
spectra of both descriptions agree since the BPS part of supergravity --- the first sum in (\ref{multiplet_spectrum}) ---
agrees exactly with (\ref{BPSstring}). Here we have used (\ref{largeN4BPS}), namely that each BPS
representation of the large ${\cal N}=4$ superconformal algebra contains two BPS representations of 
the supergravity symmetry algebra $D(2,1|\alpha)$. 

Our finding resolves a number of long-standing puzzles. In particular, the mysterious fact that the BPS
bound for $D(2,1|\alpha)$ is strictly weaker than that of the large ${\cal N}=4$ superconformal algebra --- compare
(\ref{A30}) and (\ref{A28}) --- seemed to imply that the supergravity BPS states have to acquire 
miraculous quantum corrections \cite{de Boer:1999rh,Gukov:2004ym} in order to even satisfy the stringy
BPS bound. This problem has now disappeared since the bounds only differ for $j^+\neq j^-$ --- but for 
that case, there simply aren't any supergravity BPS states (and the supergravity states that do exist
already satisfy the stringy BPS bound without any correction).

The other important consequence is that our result changes significantly the expectations for what the CFT
dual of string theory on AdS$_3\times {\rm S}^3 \times {\rm S}^3 \times {\rm S}^1$ should be. In particular,
the symmetric orbifold of the ${\cal S}_0$ theory that was first proposed for the case of $k^+=k^-$ in \cite{Elitzur:1998mm}
and essentially ruled out because of its failure to reproduce the alleged BPS spectrum of supergravity 
\cite{Gukov:2004ym} now appears to be a viable candidate after all; we will come back to analysing this 
possibility in more detail elsewhere \cite{inprep}. We should also mention that, using integrability
techniques, the BPS spectrum of string theory was analysed in \cite{BOSST}, and that also from that perspective
only BPS states with $j^+ = j^-$ were found. This suggests that  the BPS spectrum agrees for all (generic)
points in moduli space  (as is also the case for ${\rm AdS}_3 \times {\rm S}^3 \times {\cal M}_4$ with
${\cal M}_4=\mathbb{T}^4$ or ${\cal M}={\rm K3}$); thus one should be able to identify the dual CFT
directly based on the BPS spectrum, without any need to resort to the index techniques of 
\cite{Gukov:2004fh}.

\section*{Acknowledgements}

This paper is largely based on the Master thesis of one of us (LE). 
We thank Alessandro Sfondrini for comments on the draft and for sharing with us \cite{BOSST}
prior to publication.  We also thank Jan de Boer and Greg Moore for email correspondences,
and Kevin Ferreira and Juan Jottar for discussions.
The work of MRG is 
partly supported by the NCCR SwissMAP, funded by the Swiss National Science Foundation. 
He also thanks ITP of the Chinese Academy of Science for hospitality during the final stages
of this work.

\appendix

\section{BPS representations of $D(2,1|\alpha)$ and the ${\cal N}=4$ algebra}\label{app:superalgebra}

\subsection{The $D(2,1|\alpha)$ algebra}\label{app:D21}

The superalgebra $D(2,1|\alpha)$ is generated by 
\be\label{D21alpha}
L_0\ , L_{\pm 1} \ , \quad G^a_{\pm \frac{1}{2}} \ , \qquad A^{\pm, i}_{0} \ . 
\ee
Here $a\in\{0,1,2,3\}$ and $i\in\{1,2,3\}$, and the commutation relations are 
\begin{eqnarray}
{}[L_m,L_n] & = & (m-n)\, L_{m+n} \\
{}[L_m, G^a_{r}] & = & (\tfrac{m}{2} - r) G^a_{m+r} \\
{}[A^{\pm, i}_{0}, G^a_r] & = & {\rm i}\, \alpha^{\pm\, i}_{ab} G^b_r  \label{B6}\\
{}[A^{\pm , i}_0,A^{\pm, j}_0] & = & {\rm i} \, \epsilon^{ijl} A^{\pm, l}_0 \\
{}\{G^a_r, G^b_s\} & = & 2 \delta^{ab} \, L_{r+s}
+ 4\, (r-s)\, \left(\gamma \,  {\rm i}\, \alpha^{+\, i}_{ab}\, A^{+, i}_{r+s} + (1-\gamma) \, {\rm i}\, \alpha^{-\, i}_{ab}\, A^{-, i}_{r+s} \right) \ ,
\end{eqnarray}
while $[L_m,A^{\pm, i}_0]=0$. Furthermore, the expressions $\alpha^{\pm\, i}_{ab}$ are the $4\times 4$ matrices
\begin{equation}
\alpha^{\pm\, i}_{ab} = \frac{1}{2} \Bigl( \pm \delta_{ia} \delta_{b0} \mp \delta_{ib} \delta_{a0} + \epsilon_{iab} \Bigr) \ ,
\end{equation}
that satisfy the relations 
\begin{equation}
{}[\alpha^{\pm\, i},\alpha^{\pm\, j}] = -\epsilon^{ijl}\, \alpha^{\pm\, l} \ , \qquad
{}[\alpha^{+\, i},\alpha^{-\, j}] = 0 \ , \qquad
{}\{\alpha^{\pm\, i},\alpha^{\pm\, j} \} = - \tfrac{1}{2}\, \delta^{ij} \ .
\end{equation}
The parameter $\gamma$ that appears in these commutation relations is expressed in terms of $\alpha$ as 
\be
\gamma = \frac{\alpha}{1+\alpha} \ , 
\ee
or equivalently 
\be\label{alphgam}
\alpha = \frac{\gamma}{1-\gamma} \ . 
\ee
Note that the algebra is isomorphic under $\gamma\leftrightarrow (1-\gamma)$; in terms of $\alpha$ this
is the transformation $\alpha\leftrightarrow \alpha^{-1}$.

\subsubsection{BPS representations}\label{app:D21reps}

The highest weight representations of $D(2,1|\alpha)$ are labelled by $j^+, j^-, h$, where $j^\pm$ are
the spins of the two $\mathfrak{su}(2)$ algebras generated by $A^{\pm\, i}_0$, while $h$ is the eigenvalue of $L_0$. 
(The highest weight states are annihilated by the positive modes, $G^a_{1/2}$ and $L_1$.) A generic (long) representation
has the form 
\be\label{LongD21}
\begin{array}{lccccc}
& & & (j^+,j^-)  & \\
&(j^++\tfrac{1}{2},j^-+\tfrac{1}{2}) &
 (j^++\tfrac{1}{2},j^-- \tfrac{1}{2}) & (j^+-\tfrac{1}{2},j^- + \tfrac{1}{2})  & (j^+-\tfrac{1}{2},j^--\tfrac{1}{2}) \\
& (j^++1,j^-) & (j^+,j^-+1) & 2 \cdot (j^+,j^-) &  (j^+-1,j^-) & (j^+,j^--1)  \\
&(j^++\tfrac{1}{2},j^-+\tfrac{1}{2}) &
 (j^++\tfrac{1}{2},j^-- \tfrac{1}{2}) & (j^+-\tfrac{1}{2},j^- + \tfrac{1}{2})  & (j^+-\tfrac{1}{2},j^--\tfrac{1}{2}) \\
&  & & (j^+,j^-) \ , & 
\end{array}
\ee
where the different lines correspond to states with conformal dimension $h=h_0$, $h=h_0+\frac{1}{2}$, $h=h_0+1$, 
$h=h_0+\frac{3}{2}$ and $h=h_0+2$, respectively, whose $\mathfrak{su}(2)\oplus\mathfrak{su}(2)$  representation is given. 
The BPS bound takes the form, see e.g., \cite{de Boer:1999rh,Gukov:2004ym}
\begin{equation}\label{A30}
h \geq \Bigl[ \frac{1}{1+\alpha}\, j^- + \frac{\alpha}{1+\alpha} \, j^+ \Bigr]  \ .
\end{equation}
The corresponding BPS representation then consists of the subset of states, see \cite{de Boer:1999rh} eq.~(4.2) 
\be\label{BPSD21}
\begin{array}{lccc}
h=h_0 & & (j^+,j^-)  & \\[2pt]
h=h_0+\tfrac{1}{2} \hspace*{0.5cm}& (j^++\tfrac{1}{2},j^-- \tfrac{1}{2}) & (j^+-\tfrac{1}{2},j^-+\tfrac{1}{2})  & (j^+-\tfrac{1}{2},j^- -\tfrac{1}{2}) \\[2pt]
h=h_0+1  \hspace*{0.5cm} & (j^+,j^--1) & (j^+-1,j^-) & (j^+,j^-) \\[2pt]
h=h_0+\tfrac{3}{2} &  & (j^+-\tfrac{1}{2},j^--\tfrac{1}{2}) & 
\end{array}
\ee
Here $h_0=(\frac{1}{1+\alpha}\, j^- + \frac{\alpha}{1+\alpha} \, j^+)$ saturates the BPS bound. We shall denote the long
representation (\ref{LongD21}) as $[j^+,j^-]$, and the short representation (\ref{BPSD21}) as $[j^+,j^-]_s$. Note that
each long representation contains the set of states corresponding to two short representations
\be\label{shortlong}
[j^+,j^-] \cong [j^+,j^-]_s \oplus [j^++\tfrac{1}{2},j^- +\tfrac{1}{2}]_s \ . 
\ee
The above description is only correct if $j^\pm\geq 1$; for small values of $j^\pm$ the
representations are further shortened; explicit formulae for these representations are given in \cite[eq.~(4.3)]{de Boer:1999rh}.

\subsection{The large ${\cal N}=4$ superconformal algebra}\label{app:Agamma}

The large ${\cal N}=4$ superconformal algebra $A_\gamma$ whose wedge algebra is $D(2,1|\alpha)$ is defined by (we follow the 
conventions of \cite{Gaberdiel:2013vva}),
\begin{eqnarray}
{}[U_m,U_n] & = &  \tfrac{k^+ + k^-}{2} \, m \, \delta_{m,-n}  \label{A1} \\
{}[A^{\pm, i}_m, Q^a_r] & = & {\rm i}\, \alpha^{\pm\, i}_{ab} \, Q^b_{m+r}   \label{A2} \\
{} \{Q^a_r,Q^b_s \} & = &  \tfrac{k^+ + k^-}{2} \, \, \delta^{ab} \, \delta_{r,-s}  \label{A3} \\
{}[A^{\pm, i}_{m}, A^{\pm, j}_{n} ] & = &  \tfrac{k^\pm}{2}\, m \, \delta^{ij}\, \delta_{m,-n} 
+ {\rm i}\, \epsilon^{ijl}\, A^{\pm, l}_{m+n}  \label{A4} \\
{} [U_m,G^a_r] & = & m \, Q^a_{m+r}  \label{A5} \\
{}[A^{\pm, i}_{m},G^a_r] & = &  {\rm i} \,\alpha^{\pm\, i}_{ab} \,G^b_{m+r}  
\mp \tfrac{2 k^\pm }{k^++k^-}\, m \, \alpha^{\pm\, i}_{ab}\, Q^b_{m+r}  \label{A6} \\
{} \{Q^a_r,G^b_s\} & = & 2\,  \alpha^{+\, i}_{ab}\, A^{+, i}_{r+s} - 2\,  \alpha^{-\, i}_{ab}\, A^{-, i}_{r+s} + \delta^{ab} \,
U_{r+s}  \label{A7} \\
{} \{G^a_r,G^b_s\} & = & \tfrac{c}{3}\, \delta^{ab}\, (r^2 - \tfrac{1}{4}) \delta_{r,-s} 
+ 2\, \delta^{ab}\, L_{r+s} \nonumber \\
& & \ + 4\, (r-s)\, \left(\gamma \,  {\rm i}\, \alpha^{+\, i}_{ab}\, A^{+, i}_{r+s} + (1-\gamma) \, {\rm i}\, \alpha^{-\, i}_{ab}\, A^{-, i}_{r+s} \right) \ .
 \label{A8}
\end{eqnarray}
In terms of the levels of the two $\mathfrak{su}(2)$ algebras, we have
\begin{equation}\label{A9}
\gamma = \frac{k^-}{k^+ + k^-} \ , \qquad c = \frac{6 k^+ k^-}{k^+ + k^-} \ .
\end{equation}

\subsubsection{The BPS Bound}\label{app:bound}

The highest weight representations of the large superconformal ${\cal N}=4$ algebra $A_\gamma$ are characterised by 
$(h,j^\pm,u)$, where $h$ is the conformal dimension of the highest weight states, while $j^\pm$ are the spins of the two affine 
$\mathfrak{su}(2)$ algebras, and $u$ denotes the $\mathfrak{u}(1)$-charge, i.e.\ the eigenvalue under $U_0$. If we require unitarity,
we need that  $j^\pm \leq k^\pm/2$.  However, as explained in \cite{Gunaydin:1988re}, unitarity actually requires that 
\begin{equation}\label{A.23}
j^\pm \leq \frac{(k^\pm -1)}{2} \ .
\end{equation}
The BPS bound takes the form \cite{Gunaydin:1988re,Petersen:1989zz,Petersen:1989pp}
\begin{equation}\label{A28}
h \geq \frac{1}{k^++k^-} \, \Bigl[ k^+ j^- + k^- j^+ + u^2 + (j^+-j^-)^2 \Bigr] \ .
\end{equation}
Note that this bound differs from the the corresponding BPS bound of the wedge algebra 
$D(2,1|\alpha)$, see (\ref{A30}); apart from the additional $u^2$ term there is in particular also the
$(j^+-j^-)^2$ term. If we denote the corresponding representation by $[j^+,j^-,u]$ then it only satisfies
the BPS bound of $D(2,1|\alpha)$ if $u=0$ and $j^+=j^-$. On the other hand, if this is the case, the 
BPS representation $[j^+,j^-,u]$ of the linear $A_\gamma$ algebra contains actually two BPS 
representations of $D(2,1|\alpha)$ 
\be\label{largeN4BPS}
[j,j,u=0]_S  = [j,j]_s \oplus [j +\tfrac{1}{2},j \oplus \tfrac{1}{2}]_s \oplus \ \hbox{non-BPS reps of $D(2,1|\alpha)$}\ .
\ee
This is basically a consequence of the fact that in addition to the four supercharges (that also appear in 
$D(2,1|\alpha)$), $A_\gamma$ also contains four free fermions.

\section{The supergravity analysis}\label{app:sugra}

In this section we give some more details of the supergravity analysis of section~\ref{sec:sugra}. To start with, let us 
collect various identities that describe the action of differential operators on the spherical harmonics:
\begin{align}
\Delta_+ Y^{(\ell\, 0)}_+&=(1-(\ell+1)^2) Y^{(\ell\, 0)}_+\ , \label{harmonic_function_scalars}\\
\Delta_+ Y_{+,a}^{(\ell\, \pm 1)}&=(2-(\ell+1)^2) Y^{(\ell\, \pm 1)}_{+,a}\ , \\
\nabla^a Y_{+,a}^{(\ell\, \pm 1)}&=0\ , \\
\Delta_+ Y_{+,ab}^{(\ell\, \pm 2)}&= (3-(\ell+1)^2) Y^{(\ell\, \pm 2)}_{+,ab}\ , \\
\nabla^a Y_{+,ab}^{(\ell\, \pm 2)} &=0\ , \\
g^{ab} Y_{+,ab}^{(\ell\, \pm 2)}&=0\ , \\
\tensor{\epsilon}{_a^{bc}}\partial_b Y_{+,c}^{(\ell\, \pm 1)}&= \pm (\ell+1) Y_{+,a}^{(\ell\, \pm 1)}\ .
\end{align}
Here, $\Delta_+$ is the Laplace operator on $\mathrm{S}^3_+$. Similar formulae hold of course for $\mathrm{S}^3_-$. 

\subsection{Harmonic Expansion on $\mathrm{S}^3_+ \times \mathrm{S}^3_-$}\label{app:harmonics}
We expand the fields into harmonics as follows:
\begin{align}
H_{\mu\nu} &=\sum\nolimits_{\ell^+,\ell^-} H_{\mu\nu}^{(\ell^+\, 0)(\ell^-\, 0)} Y_+^{(\ell^+\, 0)}Y_-^{(\ell^-\, 0)}\ , \label{harmonic_expansion_H}\\
\phi&=\sum\nolimits_{\ell^+,\ell^-} \phi^{(\ell^+\, 0) (\ell^-\, 0)} Y_+^{(\ell^+\, 0)}Y_-^{(\ell^-\, 0)}\ , \\
M&=\sum\nolimits_{\ell^+,\ell^-} M^{(\ell^+\, 0) (\ell^-\, 0)} Y_+^{(\ell^+\, 0)}Y_-^{(\ell^-\, 0)}\ , \\
N&=\sum\nolimits_{\ell^+,\ell^-} N^{(\ell^+\, 0) (\ell^-\, 0)} Y_+^{(\ell^+\, 0)}Y_-^{(\ell^-\, 0)}\ , \\
P&=\sum\nolimits_{\ell^+,\ell^-} P^{(\ell^+\, 0) (\ell^-\, 0)} Y_+^{(\ell^+\, 0)}Y_-^{(\ell^-\, 0)}\ , \\
U_\rho&=\sum\nolimits_{\ell^+,\ell^-} U_\rho^{(\ell^+\, 0) (\ell^-\, 0)} Y_+^{(\ell^+\, 0)}Y_-^{(\ell^-\, 0)}\ ,\\
V_{a}&=\sum\nolimits_{\ell^+,\ell^-} V^{(\ell^+\, \pm 1) (\ell^-\, 0)} Y_{+,a}^{(\ell^+\, \pm 1)}Y_-^{(\ell^-\, 0)}+V^{(\ell^+\, 0)(\ell^-\, 0)} \partial_a Y_+^{(\ell^+\, 0)}Y_-^{(\ell^-\, 0)}\ , \\
R_{\mu a}&=\sum\nolimits_{\ell^+,\ell^-} R_\mu^{(\ell^+\, \pm 1) (\ell^-\, 0)} Y_{+,a}^{(\ell^+\, \pm 1)}Y_-^{(\ell^-\, 0)}+R_{\mu}^{(\ell^+\, 0)(\ell^-\, 0)} \partial_a Y_+^{(\ell^+\, 0)}Y_-^{(\ell^-\, 0)}\ , \\
C_{\mu a}&=\sum\nolimits_{\ell^+,\ell^-} C_\mu^{(\ell^+\, \pm 1) (\ell^-\, 0)} Y_{+,a}^{(\ell^+\, \pm 1)}Y_-^{(\ell^-\, 0)}+C_{\mu}^{(\ell^+\, 0)(\ell^-\, 0)} \partial_a Y_+^{(\ell^+\, 0)}Y_-^{(\ell^-\, 0)}\ , \\
W_{i}&=\sum\nolimits_{\ell^+,\ell^-} W^{(\ell^+\, 0) (\ell^-\, \pm 1)} Y_+^{(\ell^+\, 0)}Y_{-,i}^{(\ell^-\, \pm 1)}+W^{(\ell^+\, 0)(\ell^-\, 0)} Y_+^{(\ell^+\, 0)}\partial_i Y_-^{(\ell^-\, 0)}\ , \\
S_{\mu i}&=\sum\nolimits_{\ell^+,\ell^-} S_\mu^{(\ell^+\, 0) (\ell^-\, \pm 1)} Y_+^{(\ell^+\, 0)}Y_{-,i}^{(\ell^-\, \pm 1)}+S_{\mu}^{(\ell^+\, 0)(\ell^-\, 0)} Y_+^{(\ell^+\, 0)}\partial_i Y_-^{(\ell^-\, 0)}\ , \\
D_{\mu i}&=\sum\nolimits_{\ell^+,\ell^-} D_\mu^{(\ell^+\, 0) (\ell^-\, \pm 1)} Y_+^{(\ell^+\, 0)}Y_{-,i}^{(\ell^-\, \pm 1)}+D_{\mu}^{(\ell^+\, 0)(\ell^-\, 0)} Y_+^{(\ell^+\, 0)}\partial_i Y_-^{(\ell^-\, 0)}\ , \\
T_{a i}&=\sum\nolimits_{\ell^+,\ell^-} T^{(\ell^+\, \pm 1) (\ell^-\, \pm 1)} Y_{+,a}^{(\ell^+\, \pm 1)}Y_{-,i}^{(\ell^-\, \pm 1)}+T^{(\ell^+\, 0)(\ell^-\, \pm 1)} \partial_a Y_+^{(\ell^+\, 0)} Y_{-,i}^{(\ell^-\, \pm 1)}\nonumber\\
&\qquad\qquad+T^{(\ell^+\, \pm 1)(\ell^-\, 0)}Y_{+,a}^{(\ell^+\, \pm 1)} \partial_i Y_-^{(\ell^-\, 0)}+T^{(\ell^+\, 0)(\ell^-\, 0)}\partial_a Y_+^{(\ell^+\, 0)} \partial_i Y_-^{(\ell^-\, 0)}\ , \\
E_{a i}&=\sum\nolimits_{\ell^+,\ell^-} E^{(\ell^+\, \pm 1) (\ell^-\, \pm 1)} Y_{+,a}^{(\ell^+\, \pm 1)}Y_{-,i}^{(\ell^-\, \pm 1)}+E^{(\ell^+\, 0)(\ell^-\, \pm 1)} \partial_a Y_+^{(\ell^+\, 0)} Y_{-,i}^{(\ell^-\, \pm 1)}\nonumber\\
&\qquad\qquad+E^{(\ell^+\, \pm 1)(\ell^-\, 0)}Y_{+,a}^{(\ell^+\, \pm 1)} \partial_i Y_-^{(\ell^-\, 0)}+E^{(\ell^+\, 0)(\ell^-\, 0)}\partial_a Y_+^{(\ell^+\, 0)} \partial_i Y_-^{(\ell^-\, 0)}\ , \\
K_{ab}&=\sum\nolimits_{\ell^+,\ell^-} K^{(\ell^+\, \pm 2)(\ell^-\, 0)} Y_{+,ab}^{(\ell^+\, \pm 2)}Y_-^{(\ell^-\, 0)}+K^{(\ell^+\, \pm 1)(\ell^-\, 0)} \nabla_{\{a} Y_{+,b\}}^{(\ell^+\, \pm 1)}Y_-^{(\ell^-\, 0)}\nonumber\\
&\qquad\qquad+K^{(\ell^+\, 0)(\ell^-\, 0)}\nabla_{\{a}\nabla_{b\}} Y_+^{(\ell^+\, 0)}Y_-^{(\ell^-\, 0)}\ , \\
L_{ij}&=\sum\nolimits_{\ell^+,\ell^-} L^{(\ell^+\, 0)(\ell^-\, \pm 2)} Y_{+}^{(\ell^+\, 0)}Y_{-,ij}^{(\ell^-\, \pm 2)}+K^{(\ell^+\, 0)(\ell^-\, \pm 1)} Y_+^{(\ell^+\, 0)}\nabla_{\{i} Y_{-,j\}}^{(\ell^-\, \pm 1)}\nonumber\\
&\qquad\qquad+L^{(\ell^+\, 0)(\ell^-\, 0)} Y_+^{(\ell^+\, 0)}\nabla_{\{i}\nabla_{j\}}Y_-^{(\ell^-\, 0)}\ , \label{harmonic_expansion_L}
\end{align}
i.e.\ for each $\ell^\pm \in \mathbb{N}_0$ we have one set of harmonics.

\subsection{Splitting the Equations of Motion}

In this section we study the equations of motion for the fluctuations 
\eqref{harmonic_expansion_H} -- \eqref{harmonic_expansion_L}. Inserting them into the quadratic action \eqref{action_fluctuations_simplified}
we find the following equations of motion:\\
{\bf Dilaton:}
\begin{align}
0&=-(\Delta_0+\Delta_++\Delta_-)\phi+\frac{1}{4}\Delta_0(2M+3N+3P)+\frac{1}{4}\Delta_+(3M+2N+3P)\nonumber\\
&\qquad+\frac{1}{4}\Delta_-(3M+3N+2P)-\frac{1}{2}r^{-1}\nabla_\lambda U^\lambda+\frac{1}{2}r_+^{-1}\nabla_a V^a+\frac{1}{2}r_-^{-1}\nabla_i W^i\nonumber\\
&\qquad-\frac{1}{4}\nabla_\mu \nabla_\nu H^{\mu\nu}-\frac{1}{4}\nabla_i\nabla_j L^{ij}-\frac{1}{2}\nabla_\mu \nabla_i S^{\mu i}\ .\label{dilaton}
\end{align}
{\bf Metric:} \\
$\mu\nu$-trace component:
\begin{align}
0&=-(\Delta_0+3\Delta_++3\Delta_--12r^{-2})M-\Delta_0(3N+3P-4\phi)-\Delta_+\left(3N+\frac{9}{2}P-6\phi\right)\nonumber\\
&\qquad-\Delta_-\left(\frac{9}{2}N+3P-6\phi\right)-3r^{-1}\nabla_\mu U^\mu -3r_+^{-1}\nabla_a V^a-3 r_-^{-1}\nabla_i W^i\nonumber\\
&\qquad+\frac{1}{2}\nabla_\mu \nabla_\nu H^{\mu\nu}+\frac{3}{2}\nabla_i\nabla_j L^{ij}+2\nabla_\mu \nabla_i S^{\mu i}\ .\label{metricmunutrace}
\end{align}
$\mu\nu$-traceless component:
\begin{align}
0&=-(\Delta_0+\Delta_++\Delta_--4r^{-2})H_{\mu\nu}+2\nabla_\rho \nabla_{\{\mu}\tensor{H}{_{\nu\}}^\rho}+2\nabla_i\nabla_{\{\mu}\tensor{S}{_{\nu\}}^i}\nonumber\\
&\qquad-\nabla_{\{\mu}\nabla_{\nu\}} (M+3N+3P-4\phi)\ .\label{metricmunutraceless}
\end{align}
$ab$-trace component:
\begin{align}
0&=-\left(\Delta_0+\frac{1}{3}\Delta_++\Delta_-+4 r_+^{-2}\right)N-\Delta_0\left(M+\frac{3}{2}P-2\phi\right)-\Delta_+\left(M+P-\frac{4}{3}\phi\right)\nonumber\\
&\qquad-\Delta_-\left(\frac{3}{2}M+P-2\phi\right)+r^{-1}\nabla_\lambda U^\lambda+r_+^{-1}\nabla_a V^a-r_-^{-1}\nabla_i W^i\nonumber\\
&\qquad+\frac{1}{2}\nabla_\mu \nabla_\nu H^{\mu\nu}+\frac{1}{2}\nabla_i\nabla_j L^{ij}+\nabla_\mu\nabla_i S^{\mu i}\ .\label{metricabtrace}
\end{align}
$ab$-traceless component:
\begin{align}
0&=-(\Delta_0+\Delta_++\Delta_--2r_+^{-2})K_{ab}-\nabla_{\{a}\nabla_{b\}}(3M+N+3P-4\phi)\nonumber\\
&\qquad+2\nabla_\mu \nabla_{\{a}\tensor{R}{^\mu_{b\}}}+2\nabla_i \nabla_{\{a}\tensor{T}{_{b\}}^i}\ .\label{metricabtraceless}
\end{align}
$ij$-trace component:
\begin{align}
0&=-\left(\Delta_0+\Delta_++\frac{1}{3}\Delta_-+4r_-^{-2}\right)P-\Delta_0\left(M+\frac{3}{2}N-2\phi\right)\nonumber\\
&\qquad-\Delta_+\left(\frac{3}{2}M+N-2\phi\right)-\Delta_-\left(M+N-\frac{4}{3}\phi\right)+r^{-1}\nabla_\lambda U^\lambda-r_+^{-1}\nabla_a V^a\nonumber\\
&\qquad+r_-^{-1}\nabla_i W^i+\frac{1}{2}\nabla_\mu \nabla_\nu H^{\mu\nu}+\frac{1}{6}\nabla_i\nabla_j L^{ij}+\frac{2}{3}\nabla_\mu\nabla_i S^{\mu i} .\label{metricijtrace}
\end{align}
$ij$-traceless component:
\begin{align}
0&=-(\Delta_0+\Delta_++\Delta_-+4r_-^{-2})L_{ij}-\nabla_{\{i}\nabla_{j\}}(3M+3N+P-4\phi)\nonumber\\
&\qquad +2\nabla_k\nabla_{\{i}\tensor{L}{_{j\}}^k}+2\nabla_\mu\nabla_{\{i}\tensor{S}{^\mu_{j\}}}\ .\label{metricijtraceless}
\end{align}
$\mu a$-component:
\begin{align}
0&=-\left(\Delta_0+\Delta_++\Delta_--2r_+^{-2}\right)R_{\mu a}+\nabla_\lambda\nabla_\mu \tensor{R}{^\lambda_a}
+\nabla_\lambda\nabla_a \tensor{H}{_\mu^\lambda}+\nabla_i\nabla_\mu \tensor{T}{_a^i}+\nabla_i\nabla_a \tensor{S}{_\mu^i}\nonumber\\
&\qquad+\nabla_b\nabla_\mu \tensor{K}{_a^b}-\nabla_\mu \nabla_a \left(2M+2N+3P-4\phi\right)
+2r^{-1}\nabla_a U_\mu -2r_+^{-1}\nabla_\mu V_a\nonumber\\
&\qquad-2r_+^{-1}\epsilon_{abc}\nabla^c \tensor{C}{_\mu^b}+2r^{-1}\epsilon_{\mu \lambda\nu}\nabla^\nu \tensor{C}{^\lambda_a}\ . \label{metricmua}
\end{align}
$\mu i$-component:
\begin{align}
0&=-\left(\Delta_0+\Delta_++\Delta_-\right)S_{\mu i}+\nabla_\lambda\nabla_\mu \tensor{S}{^\lambda_i}
+\nabla_j\nabla_\mu\tensor{L}{_i^j}+\nabla_\lambda\nabla_i\tensor{H}{_\mu^\lambda}+\nabla_j\nabla_i \tensor{S}{_\mu^j}\nonumber\\
&\qquad-\nabla_\mu\nabla_i\left(2M+3N+2P-4\phi\right)+2r^{-1}\nabla_i U_\mu -2r_-^{-1}\nabla_\mu W_i\nonumber\\
&\qquad-2r_-^{-1}\epsilon_{ijk}\nabla^k\tensor{D}{_\mu^j}+2r^{-1}\epsilon_{\mu\lambda\nu}\nabla^\nu \tensor{D}{^\lambda_i}\ . \label{metricmui}
\end{align}
$ai$-component:
\begin{align}
0&=-\left(\Delta_0+\Delta_++\Delta_--2r_+^{-2}\right)T_{ai}+\nabla_j\nabla_a\tensor{L}{_i^j}+\nabla_\mu\nabla_i\tensor{R}{^\mu_a}+\nabla_\mu\nabla_a\tensor{S}{^\mu_i}+\nabla_j\nabla_i\tensor{T}{_a^j}\nonumber\\
&\qquad-\nabla_i\nabla_a\left(3M+2N+2P-4\phi\right)-2r_-^{-1}\nabla_a W_i-2r_+^{-1}\nabla_i V_a\nonumber\\
&\qquad+2r_+^{-1}\epsilon_{abc}\nabla^c\tensor{E}{^b_i}-2r_-^{-1}\epsilon_{ijk}\nabla^k\tensor{E}{_a^j}\ .\label{metricai}
\end{align}
{\bf Kalb-Ramond field:}\\
$\mu\nu$-component (contracted with $\epsilon^{\mu\nu\lambda}$):
\begin{align}
0&=-\nabla^\lambda\nabla_\mu U^\mu -(\Delta_++\Delta_-)U^\lambda+r^{-1}\nabla^\lambda(3M-3N-3P+4\phi)\nonumber\\
&\qquad+2r^{-1}\nabla_i S^{\lambda i}-\tensor{\epsilon}{^\lambda_\mu_\nu}\nabla_i\nabla^\nu D^{\mu i}-\tensor{\epsilon}{^\lambda_\mu_\nu}\nabla_a\nabla^\nu C^{\mu a}\ .\label{kalbmu}
\end{align}
$ab$-component (contracted with $\epsilon^{abc}$):
\begin{align}
0&=-\nabla^c\nabla_a V^a-(\Delta_0+\Delta_-)V^c+r_+^{-1}\nabla^c(-3M+3N-3P+4\phi)+2r_+^{-1}\nabla_i T^{ci}\nonumber\\
&\qquad+2r_+^{-1}\nabla_\mu R^{\mu c}+\tensor{\epsilon}{^c_{ab}}\nabla_i\nabla^b E^{ai}-\tensor{\epsilon}{^c_{ab}}\nabla_\lambda\nabla^b C^{\lambda a}\ .\label{kalba}
\end{align}
$ij$-component (contracted with $\epsilon^{ijk}$):
\begin{align}
0&=-\nabla^k\nabla_i W^i-(\Delta_0+\Delta_+)W^k+r_-^{-1} \nabla^k(-3M-3N+3P+4\phi)+2r_-^{-1}\nabla_\mu S^{\mu k}\nonumber\\
&\qquad-\tensor{\epsilon}{^k_{ij}}\nabla_a\nabla^j E^{ai}-\tensor{\epsilon}{^k_{ij}}\nabla_\mu \nabla^j D^{\mu i}\ . \label{kalbi}
\end{align}
$\mu a$-component:
\begin{align}
0&=-\left(\Delta_0+\Delta_++\Delta_--2r_+^{-2}\right)C_{\mu a}+\nabla_\lambda\nabla_\mu \tensor{C}{^\lambda_a}+\nabla_i\nabla_a \tensor{D}{_\mu^i}-\nabla_\mu \nabla_i \tensor{E}{_a^i}\nonumber\\
&\qquad+2r^{-1}\epsilon_{\mu\lambda\nu}\nabla^\nu \tensor{R}{^\lambda_a}-2r_+^{-1}\epsilon_{abc}\nabla^c \tensor{R}{_\mu^b}+\epsilon_{abc}\nabla^c\nabla_\mu V^b-\epsilon_{\mu\lambda\nu}\nabla^\nu\nabla_a U^\lambda\ .\label{kalbmua}
\end{align}
$\mu i$-component:
\begin{align}
0&=-\left(\Delta_0+\Delta_++\Delta_-\right)D_{\mu i}+\nabla_\lambda\nabla_\mu \tensor{D}{^\lambda_i}+\nabla_j\nabla_i\tensor{D}{_\mu^j}\nonumber\\
&\qquad+2r^{-1}\epsilon_{\mu\lambda\nu}\nabla^\nu\tensor{S}{^\lambda_i}-2r_-^{-1}\epsilon_{ijk}\nabla^k\tensor{S}{_\mu^j}+\epsilon_{ijk}\nabla^k\nabla_\mu W^j-\epsilon_{\mu\lambda\nu}\nabla^\nu\nabla_i U^\lambda\ . \label{kalbmui}
\end{align}
$ai$-component:
\begin{align}
0&=-\left(\Delta_0+\Delta_++\Delta_--2r_+^{-2}\right)E_{ai}+\nabla_j\nabla_i\tensor{E}{_a^j}
-\nabla_\lambda\nabla_i \tensor{C}{^\lambda_a}+\nabla_\lambda\nabla_a \tensor{D}{^\lambda_i}
\nonumber\\
&\qquad+2 r_+^{-1}\epsilon_{abc}\nabla^c\tensor{T}{^b_i}-2r_-^{-1}\epsilon_{ijk} \nabla^k \tensor{T}{_a^j}+\epsilon_{ijk}\nabla^k\nabla_a W^j\ .\label{kalbai}
\end{align}
Note that we have broken the symmetry $\mathrm{S}^3_+ \leftrightarrow \mathrm{S}^3_-$ by our gauge choice, see
eq.~(\ref{gauge_conditions}). 

\subsection{Scalar Part of the Equations of Motion}
From now on, we discuss the generic case where $\ell^+\ge 2$ and $\ell^-\ge 2$. For low $\ell^\pm$, there are additional issues to be taken care of;
these cases are discussed in Appendix~\ref{app:special}.
We now extract the scalar part of these equations. We have the following scalars:
\be 
M,\ N,\ P,\ \phi,\ \nabla^\mu U_\mu,\ \nabla^a V_a,\ \nabla^i W_i,\ \nabla_\mu\nabla_\nu H^{\mu\nu} ,\ \nabla_i\nabla_j L^{ij} ,\
\nabla_\mu\nabla_i S^{\mu i} , \ \nabla_\mu \nabla_i D^{\mu i}\ .
\ee
Note that we have
\begin{align}
\nabla^a V_a&=-r_+^{-2}\sum\nolimits_{\ell^+,\ell^-}\ell^+(\ell^++2) V^{(\ell^+\, 0)(\ell^-\, 0)} Y_+^{(\ell^+\, 0)(\ell^-\, 0)}Y_-^{(\ell^+\, 0)(\ell^-\, 0)}\ ,\\
\nabla^i W_i&=-r_-^{-2}\sum\nolimits_{\ell^+,\ell^-}\ell^-(\ell^-+2) W^{(\ell^+\, 0)(\ell^-\, 0)} Y_+^{(\ell^+\, 0)(\ell^-\, 0)}Y_-^{(\ell^+\, 0)(\ell^-\, 0)}\ ,\\
\nabla^i\nabla^j L_{ij}&=\frac{2}{3}r_-^{-4}\sum\nolimits_{\ell^+,\ell^-} (\ell^--1)\ell^-(\ell^-+2)(\ell^-+3) L^{(\ell^+\, 0)(\ell^-\, 0)} 
Y_+^{(\ell^+\, 0)(\ell^-\, 0)}Y_-^{(\ell^+\, 0)(\ell^-\, 0)}\ .
\end{align}
Since in the following, only $\Omega^{(\ell^+\, 0)(\ell^-\, 0)}$ for some quantity $\Omega$ appears and we will consider fixed $\ell^+$ 
and $\ell^-$, we will use the shorthand notation
\begin{align}
\Phi&=\phi^{(\ell^+\, 0)(\ell^-\, 0)}\ , \\
\mathcal{M}&=M^{(\ell^+\, 0)(\ell^-\, 0)}\ , \\
\mathcal{N}&=N^{(\ell^+\, 0)(\ell^-\, 0)}\ , \\
\mathcal{P}&=P^{(\ell^+\, 0)(\ell^-\, 0)}\ , \\
\mathcal{U}&=r^{-1}\nabla^\mu U_\mu^{(\ell^+\, 0)(\ell^-\, 0)}\ , \\
\mathcal{H}&=\nabla^\mu\nabla^\nu H_{\mu\nu}^{(\ell^+\, 0)(\ell^-\, 0)}\ , \\
\mathcal{V}&=-r_+^{-3}\ell^+(\ell^++2) V^{(\ell^+\, 0)(\ell^-\, 0)}\ , \\
\mathcal{W}&=-r_-^{-3}\ell^-(\ell^-+2) W^{(\ell^+\, 0)(\ell^-\, 0)}\ , \\
\mathcal{S}&=-r_-^{-2}\ell^-(\ell^-+2) \nabla^\mu S_\mu ^{(\ell^+\, 0)(\ell^-\, 0)}\ , \\
\mathcal{D}&=-r_-^{-2}\ell^-(\ell^-+2) \nabla^\mu D_\mu^{(\ell^+\, 0)(\ell^-\, 0)}\ , \\
\mathcal{L}&=\frac{2}{3}r_-^{-4}(\ell^--1)\ell^-(\ell^-+2)(\ell^-+3) L^{(\ell^+\, 0)(\ell^-\, 0)}\ . \label{definition_scalar_L}
\end{align}
Let us first discuss the equations. \eqref{kalbmua} gives immediately $\mathcal{D}=0$, and this then also implies immediately 
the scalar parts of \eqref{kalbmui} and \eqref{kalbai}. On \eqref{metricabtraceless}, we will apply $\nabla^a\nabla^b$, then we will get an algebraic equation relating 
$\Phi$, $\mathcal{M}$, $\mathcal{N}$ and $\mathcal{P}$. Extracting the scalar part of \eqref{metricai} by applying $\nabla^a\nabla^i$ will yield a further algebraic equation. A last algebraic equation will come from a combination of \eqref{metricmunutrace} and \eqref{metricabtrace}. We then have four 
algebraic equations, cutting down the number of scalar fields to seven. 
These seven fields combine with the three scalar fields we found earlier to yield a total of ten scalar fields in the compactification. 

Let us begin with this outlined program. 
From \eqref{metricabtraceless}, we can upon application of $\nabla^a\nabla^b$ directly deduce that 
\be 
3\mathcal{M}+\mathcal{N}+3\mathcal{P}-4\Phi=0\ . \label{MNPphi_constraint}
\ee
So we will eliminate the field $\Phi$, furthermore $\mathcal{D}=0$ holds as already mentioned. The remaining equations written with these replacements are:\\
{\bf Dilaton:}
\begin{align}
0&=\Delta_0\left(-\frac{1}{4}\mathcal{M}+\frac{1}{2}\mathcal{N}\right)-\frac{1}{4}\left(r_+^{-2}\ell^+(\ell^++2)
+r_-^{-2}\ell^-(\ell^-+2)\right)\mathcal{N}+\frac{1}{4}r_-^{-2}\ell^-(\ell^-+2)\mathcal{P}\nonumber\\
&\qquad-\frac{1}{2}\mathcal{U}+\frac{1}{2}\mathcal{V}+\frac{1}{2}\mathcal{W}-\frac{1}{4}\mathcal{H}
-\frac{1}{2}\mathcal{S}-\frac{1}{4}\mathcal{L}\ . \label{dilaton_scalar}
\end{align}
{\bf Metric:} \\
$\mu\nu$-trace component:
\begin{align}
0&=\Delta_0(2\mathcal{M}-2\mathcal{N})+r_-^{-2}\ell^-(\ell^-+2)\left(-\frac{3}{2}\mathcal{M}+3\mathcal{N}-\frac{3}{2}\mathcal{P}\right)
+12r^{-2}\mathcal{M}\nonumber\\
&\qquad +r_+^{-2}\ell^+(\ell^++2)\left(-\frac{3}{2}\mathcal{M}+\frac{3}{2}\mathcal{N}\right)-3\mathcal{U}-3\mathcal{V}-3\mathcal{W}
+\frac{1}{2}\mathcal{H}+3\mathcal{S}-\frac{3}{2}\mathcal{L}\ . \label{metricmunutrace_scalar}
\end{align}
$\mu\nu$-traceless component:
\begin{align}
0&=\frac{4}{3}\Delta_0^2(\mathcal{M}-\mathcal{N})+\Delta_0\left(-4r^{-2}\mathcal{M}+4r^{-2}\mathcal{N}+\frac{1}{3}\mathcal{H}
+\frac{4}{3}\mathcal{S}\right)\nonumber\\
&\qquad+\left(r_+^{-2}\ell^+(\ell^++2)+r_-^{-2}\ell^-(\ell^-+2)\right)\mathcal{H}-4r^{-2}\mathcal{S}\ . \label{metricmunutraceless_scalar}
\end{align}
$ab$-trace component:
\begin{align}
0&=\frac{1}{2}\Delta_0\left(\mathcal{M}-\mathcal{N}\right)+\frac{1}{2}r_-^{-2}\ell^-(\ell^-+2)\left(\mathcal{N}-\mathcal{P}\right)
-4r_+^{-2}\mathcal{N}+\mathcal{U}+\mathcal{V}-\mathcal{W}\nonumber\\
&\qquad+\frac{1}{2}\mathcal{H}+\mathcal{S}+\frac{1}{2}\mathcal{L}\ . \label{metricabtrace_scalar}
\end{align}
$ij$-trace component:
\begin{align}
0&=\Delta_0\left(\frac{1}{2}\mathcal{M}-\mathcal{N}+\frac{1}{2}\mathcal{P}\right)+\left(\frac{1}{2}r_+^{-2}\ell^+(\ell^++2)
+\frac{2}{3}r_-^{-2}\ell^-(\ell^-+2)\right)\left(\mathcal{N}-\mathcal{P}\right)\nonumber\\
&\qquad-4r_-^{-2}\mathcal{P}+\mathcal{U}-\mathcal{V}+\mathcal{W}+\frac{1}{2}\mathcal{H}+\frac{2}{3}\mathcal{S}+\frac{1}{6}\mathcal{L}\ . \label{metricijtrace_scalar}
\end{align}
$ij$-traceless component:
\begin{align}
0&=-\Delta_0\mathcal{L}-4r_-^{-2}\mathcal{L}+\frac{4}{3}r_-^{-4}(\ell^--1)\ell^-(\ell^-+2)(\ell^-+3)\left(-\mathcal{N}
+\mathcal{P}\right)\nonumber\\
&\qquad+\left(r_+^{-2}\ell^+(\ell^++2)-\frac{1}{3}r_-^{-2}\ell^-(\ell^-+2)\right)\mathcal{L}
-\frac{4}{3}r_-^{-2}\left(\ell^-(\ell^-+2)-3\right)\mathcal{S}\ . \label{metricijtraceless_scalar}
\end{align}
$\mu a$-component:
\begin{align}
0&=\Delta_0\left(r_+^{-2}\ell^+(\ell^++2)(\mathcal{M}+\mathcal{N})-2\mathcal{V}\right)
+r_+^{-2}\ell^+(\ell^++2)(-2\mathcal{U}-\mathcal{H}-\mathcal{S})\ . \label{metricmua_scalar}
\end{align}
$\mu i$-component:
\begin{align}
0&=\Delta_0\left(r_-^{-2}\ell^-(\ell^-+2)(-\mathcal{M}+2\mathcal{N}-\mathcal{P})-2\mathcal{W}
+\mathcal{L}\right)-r_-^{-2}\ell^-(\ell^-+2)(\mathcal{H}+2\mathcal{U})\nonumber\\
&\qquad+r_+^{-2}\ell^+(\ell^++2)\mathcal{S}\ .\label{metricmui_scalar}
\end{align}
$ai$-component:
\begin{align}
0&=r_-^{-2}r_+^{-2}\ell^+(\ell^++2)\ell^-(\ell^-+2)(-\mathcal{N}+\mathcal{P})+2r_-^{-2}\ell^-(\ell^-+2)\mathcal{V}\nonumber\\
&\qquad+r_+^{-2}\ell^+(\ell^++2)(2\mathcal{W}-\mathcal{S}-\mathcal{L})\ . \label{metricai_scalar}
\end{align}
{\bf Kalb-Ramond field:} \\
$\mu\nu$-component:
\begin{align}
0&=\Delta_0\left(6\ell^{-1}\mathcal{M}-2\ell^{-1}\mathcal{N}-\mathcal{U}\right)+\left(r_+^{-2}\ell^+(\ell^++2)
+r_-^{-2}\ell^-(\ell^-+2)\right)\mathcal{U}+2r^{-1}\mathcal{S}\ . \label{kalbmu_scalar}
\end{align}
$ab$-component:
\begin{align}
0&=-\Delta_0\mathcal{V}-4r_+^{-2}\ell^+(\ell^++2)\mathcal{N}+\left(r_+^{-2}\ell^+(\ell^++2)
+r_-^{-2}\ell^-(\ell^-+2)\right)\mathcal{V}\ . \label{kalba_scalar}
\end{align}
$ij$-component:
\begin{align}
0&=-\Delta_0\mathcal{W}+r_-^{-2}\ell^-(\ell^-+2)\left(2\mathcal{N}-6\mathcal{P}\right)\nonumber\\
&\qquad+\left(r_+^{-2}\ell^+(\ell^++2)
+r_-^{-2}\ell^-(\ell^-+2)\right)\mathcal{W}+2\mathcal{S}\ . \label{kalbi_scalar}
\end{align}
From these equations, \eqref{metricmunutrace_scalar}$-4\times$\eqref{metricabtrace_scalar} and \eqref{metricai_scalar} 
are algebraic and hence impose algebraic relationships among the fields. We use these relations to eliminate the fields 
$\mathcal{S}$ and $\mathcal{L}$ from the equations. Once these relations are imposed, we have 
\begin{align}
0&=\frac{1}{2}\ell^+(\ell^++2) \times\eqref{metricmunutrace_scalar}-2\times\eqref{kalba_scalar}+r_+^2 \times \eqref{metricmua_scalar}\ ,\label{first_linear_dependency}\\
0&=2r_-^{-2}\ell^-(\ell^-+2)\times \eqref{metricijtrace_scalar}+\eqref{metricijtraceless_scalar}+\eqref{metricmui_scalar}
-2r_-^{-2}\times\eqref{kalbi_scalar}\ .  \label{second_linear_dependency}
\end{align}
Thus, we do not have to consider the equations \eqref{metricmua_scalar} and \eqref{kalbi_scalar} any longer. 

We will not use the equation \eqref{metricmunutraceless_scalar}, since it contains squares of Laplacians. We will however check 
that the solution we obtain fulfills also this equation. Omitting for the moment this equation, we are left with seven equations for the 
seven unknowns $\mathcal{M}$, $\mathcal{N}$, $\mathcal{P}$, $\mathcal{U}$, $\mathcal{V}$, $\mathcal{W}$, $\mathcal{H}$, 
namely \eqref{dilaton_scalar}, \eqref{metricmunutrace_scalar}, \eqref{metricijtrace_scalar}, \eqref{metricmui_scalar}, \eqref{kalbmu_scalar}, 
\eqref{kalba_scalar} and \eqref{kalbi_scalar}. For the sake of completeness, we reproduce here the solution, when solving these 
equations for the Laplacians:
\begin{align}
\Delta_0\mathcal{M}&=\left(r_+^{-2}\ell^+(\ell^++2)+r_-^{-2}\ell^-(\ell^-+2)-8r^{-2}\right)\nonumber\\
&\qquad \times\left(\mathcal{M}+4r^{-2}\mathcal{U}
+2r_-^{-2}\mathcal{W}-\frac{32}{3}r^{-2}\mathcal{H}\right)\ ,  \label{B57} \\
\Delta_0\mathcal{N}&=\frac{16}{3}r_+^{-2}\mathcal{M}+\left(r_+^{-2}\ell^+(\ell^++2)+r_-^{-2}\ell^-(\ell^-+2)\right)\mathcal{N}
+\frac{112}{3}r^{-2}r_+^{-2}\mathcal{U}\nonumber\\
&\qquad+4r_+^{-4}\ell^+(\ell^++2)\mathcal{V}+2r_-^{-2}\left(r_+^{-2}\ell^+(\ell^++2)+2\ell^-(\ell^-+2)
+\frac{32}{3}r_+^{-2}\right)\mathcal{W}\nonumber\\\
&\qquad+\frac{32}{3}r_+^{-2}\left(2r_+^{-2}\ell^+(\ell^++2)+2r_-^{-2}\ell^-(\ell^-+2)-\frac{13}{3}r_+^{-2}-\frac{37}{3}r_-^{-2}\right)\mathcal{H}\ , \\
\Delta_0 \mathcal{P}&=\left(r_+^{-2}\ell^+(\ell^++2)+r_-^{-2}\ell^-(\ell^-+2)+8r_-^{-2}\right)\mathcal{P}
-6r_-^{-4}\ell^-(\ell^-+2)\mathcal{W}\ , \\
\Delta_0 \mathcal{U}&=\frac{8}{3}\mathcal{M}+\left(r_+^{-2}\ell^+(\ell^++2)+r_-^{-2}\ell^-(\ell^-+2)
+\frac{20}{3}r^{-2}\right)\mathcal{U}-\frac{28}{3}r_-^{-2}\mathcal{W}\nonumber\\
&\qquad-\frac{8}{3}\left(r_+^{-2}\ell^+(\ell^++2)+r_-^{-2}\ell^-(\ell^-+2)+\frac{11}{3}r^{-2}\right)\mathcal{H}\ , \\
\Delta_0 \mathcal{V}&=\frac{4(r_-^2\ell^+(\ell^++2)+r_+^2\ell^-(\ell^-+2))}{3r_-^2\ell^+(\ell^++2)}\mathcal{M}+(r_+^{-2}\ell^+(\ell^++2)
+r_-^{-2}\ell^-(\ell^-+2))\mathcal{V}\nonumber\\
&\qquad-\frac{28(r_-^2\ell^+(\ell^++2)+r_+^2\ell^-(\ell^-+2))}{3r^2r_-^2\ell^+(\ell^++2)}\mathcal{U}-4\mathcal{N}
+\frac{4r_+^2\ell^-(\ell^-+2)}{3r_-^2\ell^+(\ell^++2)}\mathcal{P}\nonumber\\
&\qquad-\frac{4(7r_-^2\ell^+(\ell^++2)+r_+^2\ell^-(\ell^-+2))}{3r_-^4\ell^+(\ell^++2)}\mathcal{W}\nonumber\\
&\qquad-\frac{8}{9\ell^+(\ell^++2)r_-^4r_+^2}\Big(r_-^4\ell^+(\ell^++2)(3(\ell^+)^2+6\ell^++11)\nonumber\\
&\qquad\qquad+r_+^2r_-^2\left((6(\ell^+)^2+12\ell^+-37)\ell^-(\ell^-+2)-37\ell^+(\ell^++2)\right)\nonumber\\
&\qquad\qquad+r_+^4\ell^-(\ell^-+2)(3(\ell^-)^2+6\ell^--37)\Big)\mathcal{H}\ ,  \\
\Delta_0 \mathcal{W}&=-\frac{8}{3}\mathcal{P}+\left(r_+^{-2}\ell^+(\ell^++2)+r_-^{-2}\ell^-(\ell^-+2)\right)\mathcal{W}\ , \\
\Delta_0 \mathcal{H}&=-2\ell^{-2}\mathcal{U}-2r_-^{-2}\mathcal{W}+\left(r_+^{-2}\ell^+(\ell^++2)+r_-^{-2}\ell^-(\ell^-+2)
+\frac{28}{3}r^{-2}\right)\mathcal{H}\ . \label{B63}
\end{align}
As promised, also \eqref{metricmunutraceless_scalar} is satisfied by this solution.

\section{Special cases at low $\ell^\pm$}\label{app:special}
\subsection{Residual gauge transformations}
Before discussing the various special cases, we have to find all residual gauge transformations, since those gauge away modes with low $\ell^+$. 
There are two of those, modifying scalar fields. \\
{\bf AdS$_3\times$S$^3$-reparametrisations:}
\begin{align}
\xi_\mu&=\sum_{\ell^-} \xi_\mu^{(0\, 0)(\ell^-\, 0)} Y_-^{(\ell^-\, 0)}\ , \\
\xi_a&=0 \ , \\
\xi_i &=\sum_{\ell^-} \xi^{(0\, 0)(\ell^-\, 0)}  \nabla_i Y_-^{(\ell^-\, 0)} \ .
\end{align}
They induce the following transformations on the fields (only the non-trivial transformations are displayed):
\begin{align}
\delta H_{\mu\nu}^{(0\, 0)(\ell^-\, 0)}&=2\nabla_{\{\mu}\xi^{(0\, 0)(\ell^-\, 0)}_{\nu\}}\ , \\
\delta M^{(0\, 0)(\ell^-\, 0)}&=\frac{2}{3}\nabla^\mu \xi_\mu^{(0\, 0)(\ell^-\, 0)}\ , \\
\delta L_{ij}^{(0\, 0)(\ell^-\, 0)}&=2 \xi^{(0\, 0)(\ell^-\, 0)}\ , \\
\delta P^{(0\, 0)(\ell^-\, 0)}&=-\frac{2}{3}r_-^{-2}\ell^-(\ell^-+2)\xi^{(0\, 0)(\ell^-\, 0)} \ , \\
\delta S_\mu^{(0\, 0)(\ell^-\, 0)}&=\xi_\mu^{(0\, 0)(\ell^-\, 0)}+\nabla_\mu \xi^{(0\, 0)(\ell^-\, 0)}\ .
\end{align}
{\bf St\"uckelberg shift symmetries:}
\begin{align}
\xi_\mu&=-\sum\nolimits_{\ell^-}\nabla_\mu \lambda^{(1\, 0)(\ell^-\, 0)}Y_+^{(1\, 0)}Y_-^{(\ell^-\, 0)}\ , \\
\xi_a&=\sum\nolimits_{\ell^-}\lambda^{(1\, 0)(\ell^-\, 0)} \nabla_aY_+^{(1\, 0)}Y_-^{(\ell^-\, 0)}\ , \\
\xi_i&=-\sum\nolimits_{\ell^-}\lambda^{(1\, 0)(\ell^-\, 0)} Y_+^{(1\, 0)}\nabla_i Y_-^{(\ell^-\, 0)} \ .
\end{align}
Transforming the coordinate system with this vector field preserves the gauge \eqref{gauge_conditions} and we find for the metric
\begin{align}
\delta H_{\mu\nu}^{(1\, 0)(\ell^-\, 0)}&=-2\nabla_{\{\mu }\nabla_{\nu\}}\lambda^{(1\, 0)(\ell^-\, 0)}\ , \label{first_stuckelberg}\\
\delta M^{(1\, 0)(\ell^-\, 0)}&=-\frac{2}{3}\Delta_0 \lambda^{(1\, 0)(\ell^+\, 0)}\ , \\
\delta N^{(1\, 0)(\ell^-\, 0)}&=-2 r_+^{-2}\lambda^{(1\, 0)(\ell^+\, 0)}\ , \\
\delta P^{(1\, 0)(\ell^-\, 0)}&=\frac{2}{3}\ell^-(\ell^-+2)r_-^{-2}\lambda^{(1\, 0)(\ell^-\, 0)}\ , \\
\delta L^{(1\, 0)(\ell^-\, 0)}&=-2\lambda^{(1\, 0)(\ell^-\, 0)}\ , \\
\delta S_\mu^{(1\, 0)(\ell^-\, 0)}&=-2\nabla_\mu \lambda^{(1\, 0)(\ell^-\, 0)}  \ . \label{last_stuckelberg}
\end{align}
We will not need the transformation properties of the antisymmetric tensor field.

\subsection{Metric, Dilaton and Kalb-Ramond Field}
Let us first describe how the analysis of Appendix~\ref{app:sugra} gets modified for small $\ell^\pm$; 
the corresponding analysis for the decoupled subsystem of Section~\ref{sec:decoupled} will be described below in 
Section~\ref{app:decoup}.

\subsubsection{$\ell^+ > 1$, $\ell^-=1$}
This is the first special case. Then we have from \eqref{definition_scalar_L} that $\mathcal{L}=0$. 
We still have exactly the same algebraic relationships, i.e. $\mathcal{D}=0$, \eqref{MNPphi_constraint}, 
\eqref{metricmunutrace_scalar}$-4\times$\eqref{metricabtrace_scalar} and \eqref{metricai_scalar}. We use them to 
eliminate also $\mathcal{W}$ in this case, so we remain only with six fields. We have the same linear relationships 
as \eqref{first_linear_dependency} and \eqref{second_linear_dependency}. On top of this, \eqref{metricijtraceless_scalar} 
just vanishes trivially. Thus, in this case, we are left with six equations for the six unknowns. They can again be solved and 
\eqref{metricmunutraceless_scalar} is automatically satisfied by the solution. The eigenvalues of the corresponding matrix are 
exactly the same as before, except that now $m^2_{\ell^+,\ell^--2}$ is missing. This was to be expected, since we cannot have 
negative $\ell^-$.
\subsubsection{$\ell^+=1$, $\ell^->1$}
Now we cannot deduce any longer \eqref{MNPphi_constraint}, since the application of $\nabla^a\nabla^b$ 
on \eqref{metricabtraceless} just vanishes. However, we can use the St\"uckelberg shift symmetry 
\eqref{first_stuckelberg} -- \eqref{last_stuckelberg} 
to gauge some fields away. Using $\lambda^{(1\, 0)(\ell^-\, 0)}$ satisfying
\begin{align}
(\Delta_0+r_+^{-2}-r_-^{-2}\ell^-(\ell^-+2))\lambda^{(1\, 0)(\ell^-\, 0)}=\frac{1}{2}(3\mathcal{M}+\mathcal{N}+3\mathcal{P}-4\Phi)\ , 
 \label{stuckelberg_gauge_fix}
\end{align}
we can impose \eqref{MNPphi_constraint} as a gauge condition. Afterwards the analysis goes through as in the general case, but we 
still have a residual gauge transformation, as we can perform the St\"uckelberg shift with $\lambda^{(1\, 0)(\ell^-\, 0)}$ satisfying
\be 
(\Delta_0+r_+^{-2}-r_-^{-2}\ell^-(\ell^-+2))\lambda^{(1\, 0)(\ell^-\, 0)}=0\ .
\ee
This equation of motion coincides with the eigenvector belonging to $m^2_{\ell^+-2,\ell^-}$ and hence we can use this remaining gauge freedom 
to gauge this eigenvector away. Thus, we conclude, as in the previous case, that we obtain all fields, but $m^2_{\ell^+-2,\ell^-}$ is missing. This 
was to be expected, since the spectrum must be symmetric in $\ell^+$ and $\ell^-$.
\subsubsection{$\ell^+=\ell^-=1$}
This is just the combination of the previous cases. After setting the gauge as in  \eqref{stuckelberg_gauge_fix}, we perform the 
same analysis as for $\ell^+>1$ and $\ell^-=1$. Thus again the eigenvalue $m^2_{\ell^+,\ell^--2}$ gets removed. At the end we can use the 
residual gauge to gauge away $m^2_{\ell^+-2,\ell^-}$. Thus we are left with only five fields.
\subsubsection{$\ell^+ >1$, $\ell^-=0$}
In this case we have by definition $\mathcal{W}=\mathcal{S}=\mathcal{L}=\mathcal{D}=0$. \eqref{MNPphi_constraint} again holds 
since $\ell^+ >1$. The equations \eqref{metricijtraceless_scalar}, \eqref{metricmui_scalar}, \eqref{metricai_scalar} and \eqref{kalbi_scalar} 
are trivially satisfied, since $\nabla^i$ now annihilates everything. We are left with the six unknowns $\mathcal{M}$, $\mathcal{N}$, 
$\mathcal{P}$, $\mathcal{U}$, $\mathcal{H}$ and $\mathcal{V}$. From \eqref{metricmunutrace_scalar}$-4 \times$ \eqref{metricabtrace_scalar}, 
we get again one further algebraic constraint with which we eliminate $\mathcal{V}$. \eqref{first_linear_dependency} ist still true, the 
second linear dependency is trivial in this case. We  are thus left with five equations \eqref{dilaton_scalar}, \eqref{metricmunutrace_scalar}, 
\eqref{metricijtrace_scalar}, \eqref{kalbmu_scalar} and \eqref{kalba_scalar} for the five unknowns. The solution again passes the check 
of satisfying \eqref{metricmunutraceless_scalar}. The eigenvalues are the same as before, but this time $m^2_{\ell^+,\ell^-}$ and 
$m^2_{\ell^+,\ell^--2}$ are missing.
\subsubsection{$\ell^+=0$, $\ell^->1$}
We now have by definition $\mathcal{V}=0$. We choose $\xi_\mu^{(0\, 0)(\ell^-\, 0)}$ and $\xi^{(0\, 0)(\ell^-\, 0)}$ such that
\begin{align}
r_-^{-2}\ell^-(\ell^-+2)\xi^{(0\, 0)(\ell^-\, 0)}-\nabla^\mu \xi_\mu^{(0\, 0)(\ell^-\, 0)}&=\frac{1}{2}\left(3\mathcal{M}+\mathcal{N}+3\mathcal{P}-4\Phi\right)\ , \\
(\Delta_0-3r^{-2})\nabla^\mu \xi_\mu&=-\frac{3}{4}\mathcal{H}\ ,
\end{align}
so \eqref{MNPphi_constraint} again holds and $\mathcal{H}=0$. Now the equations \eqref{metricmua_scalar}, \eqref{metricai_scalar}, \eqref{kalba_scalar} are trivial,
furthermore the constraint $\mathcal{D}=0$ is no longer implied by \eqref{kalbmua}, \eqref{kalbmui} and \eqref{kalbai}, and their scalar part is 
now also trivial. However, $\mathcal{D}$ does not appear anywhere in the field equations and hence is not a dynamical field, so we will not 
consider it further. The residual gauge transformations 
now satisfy
\begin{align}
(\Delta_0-3r^{-2}) \nabla^\mu \xi_\mu^{(\ell^+\, 0)(\ell^-\, 0)}=0\ , \quad r_-^{-2}\ell^-(\ell^-+2)\xi^{(0\, 0)(\ell^-\, 0)}
-\nabla^\mu \xi_\mu^{(0\, 0)(\ell^-\, 0)}=0\ . \label{residual_gauge_trafos_lm1lp0} 
\end{align}
Again \eqref{metricmunutrace_scalar}$-4 \times$ \eqref{metricabtrace_scalar} is algebraic and we use it to eliminate $\mathcal{S}$. Thus, 
we remain with the fields $\mathcal{M}$, $\mathcal{N}$, $\mathcal{P}$, $\mathcal{U}$, $\mathcal{W}$, $\mathcal{L}$. 
Now \eqref{first_linear_dependency} is trivial, but \eqref{second_linear_dependency} eliminates the equation \eqref{metricijtraceless_scalar}. 
We are left with six independent equations, namely \eqref{dilaton_scalar}, \eqref{metricmunutrace_scalar}, \eqref{metricijtrace_scalar}, 
\eqref{metricmui_scalar}, \eqref{kalbmu_scalar} and \eqref{kalbi_scalar} for the six unknowns. 
The solution satisfies again \eqref{metricmunutraceless_scalar}, and the 
eigenvalues are
\be 
m^2_{\ell^++2,\ell^-}\ , \quad m^2_{\ell^+,\ell^-\pm 2} \ , \quad \lambda_{6/7}\ ,\quad  3r^{-2}\ ,
\ee
where $\lambda_{6/7}$ are as in \eqref{lambda67}. Since the equations of motion of the residual gauge transformation 
\eqref{residual_gauge_trafos_lm1lp0} and the eigenvector corresponding to $3r^{-2}$ coincide, we can gauge that eigenvector away. 
Thus, we find again the same states as in the previous subsection, except that $\ell^+$ and $\ell^-$ are interchanged.
\subsubsection{$\ell^+=0$, $\ell^-=1$}
Everything is as in the previous section, except that also $\mathcal{L}=0$ by definition. \eqref{metricijtraceless_scalar} is now 
identically zero and hence we use \eqref{second_linear_dependency} to eliminate \eqref{metricijtrace_scalar}. The resulting system 
has five unknowns, the solution fulfills again \eqref{metricmunutraceless_scalar} and the eigenvalues are
\be 
m^2_{\ell^++2,\ell^-}\ , \quad m^2_{\ell^+,\ell^-+ 2} \ , \quad \lambda_{6/7}\ ,\quad  3r^{-2}\ .
\ee
Again, using a residual gauge transformation, we can gauge the eigenvector corresponding to $3r^{-2}$ away.
\subsubsection{$\ell^+=1$, $\ell^-=0$}
We again use the St\"uckelberg transformation \eqref{stuckelberg_gauge_fix} to enforce \eqref{MNPphi_constraint}. Afterwards, 
the analysis is the same as for $\ell^-=0$ and $\ell^+>1$.  As for general $\ell^+>1$, we can gauge away the eigenvector 
corresponding to the eigenvalue $m^2_{\ell^+-2,\ell^-}$. Thus, the only remaining eigenvalues are
\be 
m^2_{\ell^++2,\ell^-}\ , \quad m^2_{\ell^+,\ell^-+ 2} \ , \quad \lambda_{6/7}\ .
\ee
\subsubsection{$\ell^+=\ell^-=0$}
We have by definition $\mathcal{V}=\mathcal{W}=\mathcal{S}=\mathcal{L}=0$. $\mathcal{D}$ again never appears in 
any equation and is hence not physical. We again perform a reparametrisation of $\mathrm{AdS}_3 \times \mathrm{S}^3_-$ to 
enforce \eqref{MNPphi_constraint}, but one easily sees that $\mathcal{H}=0$ can no longer be achieved. In fact, $\xi^{(0\, 0)(0\, 0)}$ 
never enters in the transformation. Again the combination \eqref{metricmunutrace_scalar}$-4\times$\eqref{metricabtrace_scalar} is 
algebraic and we eliminate also $\mathcal{U}$. Thus we are left with only $\mathcal{M}$, $\mathcal{N}$, $\mathcal{P}$ and $\mathcal{H}$. 
The solution again fulfills \eqref{metricmunutraceless_scalar} and has eigenvalues
\be
m^2_{\ell^+,\ell^-}\ , \quad m^2_{\ell^++2,\ell^-}\ , \quad m^2_{\ell^+,\ell^-+2}\ , \quad m^2_{\ell^++2,\ell^-+2}\ .
\ee
[Note that for $\ell^+=\ell^-=0$, $m^2_{\ell^++2,\ell^-+2} = m^2_{2,2} = \lambda_6$.]
Residual gauge transformations are $\xi_\mu^{(0\, 0)(0\, 0)}$ satisfying $\nabla^\mu \xi_\mu^{(0\, 0)(0\, 0)}=0$, i.e. setting 
\be 
\xi_\mu^{(0\, 0)(0\, 0)}=\nabla_\mu \zeta^{(0\, 0)(0\, 0)}\ ,
\ee
$\zeta^{(0\, 0)(0\, 0)}$ satisfies $\Delta_0 \zeta^{(0\, 0)(0\, 0)}=0$. Hence we can use it to gauge away 
the zero eigenvector which corresponds to $m^2_{\ell^+,\ell^-}=m^2_{0,0}=0$.

\subsection{The scalars from the decoupled subsystem}\label{app:decoup}
Finally, let us describe what happens to the scalar degrees of freedom from the decoupled subsystem, see
Section~\ref{sec:decoupled}.
We have already used up all residual gauge transformations for the metric, dilaton and the Kalb-Ramond field. We easily see 
that $\Psi$ is always present, regardless of the value of $\ell^+$ and $\ell^-$. For $\Upsilon_M$, we note that there is a residual 
gauge symmetry for $\Upsilon_M$, namely $\delta \Upsilon_M=\nabla_M \Lambda$. Choosing
\be 
\Lambda=\sum_{\ell^-} \Lambda^{(0\, 0)(\ell^-\, 0)} Y_-^{(\ell^-\, 0)}\ ,
\ee
the Lorentz gauge is preserved. If $\ell^-=0$, $\nabla^i \Omega_i=0$ and hence no scalar component of $\Upsilon_M$ survives. 
Similarly, if $\ell^+=0$, we can use this remaining gauge symmetry to set $\nabla^\mu \Pi_\mu=0$. But in this case, $\nabla^i \Omega_i$ survives, since it must not be equal to $\nabla^\mu \Pi_\mu$. It is however annihilated by $\Delta_0$ and hence can be gauged away, since we have a remaining gauge freedom of $\Lambda^{(0\, 0)(\ell^-\, 0)}$ satisfying $\Delta_0 \Lambda^{(0\, 0)(\ell^-\, 0)}=0$.

\bibliographystyle{JHEP}

\begin{thebibliography}{99}


\bibitem{David:2002wn}
J.R.~David, G.~Mandal and S.R.~Wadia,
``Microscopic formulation of black holes in string theory,''
Phys.\ Rept.\  {\bf 369} (2002) 549
{\tt [hep-th/0203048]}.

\bibitem{Gaberdiel:2007vu}
M.R.~Gaberdiel and I.~Kirsch,
``Worldsheet correlators in AdS$_3$/CFT$_2$,''
JHEP {\bf 0704} (2007) 050
{\tt [hep-th/0703001]}.

\bibitem{Dabholkar:2007ey}
A.~Dabholkar and A.~Pakman,
``Exact chiral ring of AdS$_3$ / CFT$_2$,''
Adv.\ Theor.\ Math.\ Phys.\  {\bf 13} (2009) 409
{\tt [hep-th/0703022]}.

\bibitem{deBoer:2008ss}
J.~de Boer, J.~Manschot, K.~Papadodimas and E.~Verlinde,
``The Chiral ring of AdS$_3$/CFT$_2$ and the attractor mechanism,''
JHEP {\bf 0903} (2009) 030
{\tt [arXiv:0809.0507 [hep-th]]}.

\bibitem{Gaberdiel:2014cha} 
M.R.~Gaberdiel and R.~Gopakumar,
``Higher Spins \& Strings,''
JHEP {\bf 1411} (2014) 044 
{\tt [arXiv:1406.6103 [hep-th]]}.

\bibitem{Ademollo:1976wv}
M.~Ademollo {\it et al.},
``Dual String Models with Nonabelian Color and Flavor Symmetries,''
Nucl.\ Phys.\ B {\bf 114} (1976) 297.

\bibitem{Eguchi:1987sm}
T.~Eguchi and A.~Taormina,
``Unitary Representations of $N=4$ Superconformal Algebra,''
Phys.\ Lett.\ B {\bf 196} (1987) 75.

\bibitem{Eguchi:1987wf}
T.~Eguchi and A.~Taormina,
``Character Formulas for the $N=4$ Superconformal Algebra,''
Phys.\ Lett.\ B {\bf 200} (1988) 315.

\bibitem{Gukov:2004ym}
S.~Gukov, E.~Martinec, G.~W.~Moore and A.~Strominger,
``The Search for a holographic dual to AdS$_3 \times {\rm S}^3 \times {\rm S}^3 \times {\rm S}^1$,"
Adv.\ Theor.\ Math.\ Phys.\  {\bf 9} (2005) 435
{\tt [hep-th/0403090]}.

\bibitem{Tong:2014yna}
D.~Tong,
``The holographic dual of ${\rm AdS}_{3} \times  {\rm S}^{3} \times {\rm S}^{3} \times {\rm S}^{1}$,''
JHEP {\bf 1404} (2014) 193
{\tt  [arXiv:1402.5135 [hep-th]]}.

\bibitem{Sevrin:1988ew} 
  A.~Sevrin, W.~Troost and A.~Van Proeyen,
``Superconformal algebras in two-dimensions with N=4,''
Phys.\ Lett.\ B {\bf 208} (1988) 447.

\bibitem{Schoutens:1988ig}
K.~Schoutens,
``O(n) extended superconformal field theory in superspace,''
Nucl.\ Phys.\ B {\bf 295} (1988) 634.

\bibitem{Spindel:1988sr} 
P.~Spindel, A.~Sevrin, W.~Troost and A.~Van Proeyen,
``Extended supersymmetric sigma models on group manifolds. 1. The complex structures,''
Nucl.\ Phys.\ B {\bf 308} (1988)  662.

\bibitem{VanProeyen:1989me} 
A.~Van Proeyen,
``Realizations of N=4 superconformal algebras on Wolf spaces,''
Class.\ Quant.\ Grav.\  {\bf 6} (1989) 1501.

\bibitem{Sevrin:1989ce} 
A.~Sevrin and G.~Theodoridis,
``N=4 superconformal coset theories,''
Nucl.\ Phys.\ B {\bf 332} (1990) 380.

\bibitem{Gunaydin:1988re}
M.~Gunaydin, J.L.~Petersen, A.~Taormina and A.~Van Proeyen,
``On the Unitary Representations of a Class of $N=4$ Superconformal Algebras,''
Nucl.\ Phys.\ B {\bf 322} (1989) 402.

\bibitem{Petersen:1989zz}
J.L.~Petersen and A.~Taormina,
``Characters of the $N=4$ Superconformal Algebra With Two Central Extensions,''
Nucl.\ Phys.\ B {\bf 331} (1990) 556.

\bibitem{Petersen:1989pp}
J.L.~Petersen and A.~Taormina,
``Characters of the $N=4$ Superconformal Algebra With Two Central Extensions: 2. Massless Representations,''
Nucl.\ Phys.\ B {\bf 333} (1990) 833.

\bibitem{de Boer:1999rh} 
J.~de Boer, A.~Pasquinucci and K.~Skenderis,
 ``AdS / CFT dualities involving large 2-D N=4 superconformal symmetry,''
Adv.\ Theor.\ Math.\ Phys.\  {\bf 3} (1999)  577
{\tt  [arXiv:hep-th/9904073]}.

\bibitem{Elitzur:1998mm}
S.~Elitzur, O.~Feinerman, A.~Giveon and D.~Tsabar,
``String theory on AdS$_3 \times {\rm S}^3 \times {\rm S}^3 \times {\rm S}^1$, ''
Phys.\ Lett.\ B {\bf 449} (1999) 180
{\tt [hep-th/9811245]}.

\bibitem{Maldacena:2000hw}
J.M.~Maldacena and H.~Ooguri,
``Strings in AdS(3) and SL(2,R) WZW model 1.: The Spectrum,''
J.\ Math.\ Phys.\  {\bf 42} (2001) 2929
{\tt [hep-th/0001053]}.

\bibitem{Deger:1998nm}
S.~Deger, A.~Kaya, E.~Sezgin and P.~Sundell,
``Spectrum of D = 6, N=4b supergravity on AdS in three-dimensions $\times\ {\rm S}^3$,''
Nucl.\ Phys.\ B {\bf 536} (1998) 110
{\tt [hep-th/9804166]}.

\bibitem{inprep}
L.~Eberhardt, M.R.~Gaberdiel, R.~Gopakumar and W.~Li, in progress.

\bibitem{DiVecchia:1984nyg}
P.~Di Vecchia, V.G.~Knizhnik, J.L.~Petersen and P.~Rossi,
``A Supersymmetric Wess-Zumino Lagrangian in Two-Dimensions,''
Nucl.\ Phys.\ B {\bf 253} (1985) 701.

\bibitem{Evans:1998qu}
J.M.~Evans, M.R.~Gaberdiel and M.J.~Perry,
``The no ghost theorem for AdS$_3$ and the stringy exclusion principle,''
Nucl.\ Phys.\ B {\bf 535} (1998) 152
{\tt  [hep-th/9806024]}.

\bibitem{Giveon:1998ns}
A.~Giveon, D.~Kutasov and N.~Seiberg,
``Comments on string theory on AdS$_3$,''
Adv.\ Theor.\ Math.\ Phys.\  {\bf 2} (1998) 733
{\tt [hep-th/9806194]}.

\bibitem{ortin2004gravity}
T.~Ortin, ``Gravity and strings,'' Cambridge University Press (2004). 

\bibitem{BOSST}
M.~Baggio, O.~Ohlson Sax, A.~Sfondrini, B.~Stefanski and A.~Torielli,
``Protected string spectrum in AdS$_3$/CFT$_2$ from worldsheet integrability,"
to appear.

\bibitem{Gukov:2004fh}
S.~Gukov, E.~Martinec, G.W.~Moore and A.~Strominger,
``An Index for 2-D field theories with large N = 4 superconformal symmetry,''
{\tt hep-th/0404023}.

\bibitem{Gaberdiel:2013vva} 
M.R.~Gaberdiel and R.~Gopakumar,
``Large $\mathcal{N}=4$ holography,''
{\tt arXiv:1305.4181 [hep-th]}.

\bibitem{Argurio:2000tb}
  R.~Argurio, A.~Giveon and A.~Shomer,
  ``Superstrings on AdS(3) and symmetric products,''
  JHEP {\bf 0012} (2000) 003
  {\tt [hep-th/0009242]}.

\bibitem{Gunaydin:1986fe}
M.~Gunaydin, G.~Sierra and P.K.~Townsend,
``The Unitary Supermultiplets of $d=3$ Anti-de Sitter and $d=2$ Conformal Superalgebras,''
Nucl.\ Phys.\ B {\bf 274} (1986) 429.

\bibitem{Eberhardt:2017pty}
  L.~Eberhardt, M.~R.~Gaberdiel and W.~Li,
  ``A holographic dual for string theory on $\mathrm{AdS}_3 \times \mathrm{S}^3 \times \mathrm{S}^3 \times \mathrm{S}^1$,''
  {\tt arXiv:1707.02705 [hep-th]}.



\end{thebibliography}

\end{document}